\documentclass[authoryear,10pt]{article}

\usepackage{natbib}
\usepackage{fullpage}
\usepackage{authblk}

\usepackage[utf8]{inputenc}

\usepackage{amsmath,amssymb,amsthm,amscd}
\usepackage{mathtools}
\usepackage{dsfont} % for \mathds{N} 
\usepackage{bm} % for mathematical vector
\usepackage[binary-units=true]{siunitx}
\usepackage{multirow}
\usepackage{caption}
\usepackage[font=footnotesize]{subcaption}
\usepackage{booktabs} % for \toprule \midrule \bottomrule
\usepackage{blkarray} % environment blockarray
\usepackage{pdflscape}
\usepackage{longtable}

\usepackage{tikz}
\usepackage{xcolor}
\usepackage{graphicx} 
\usepackage[colorlinks=true, citecolor=blue, linkcolor=blue]{hyperref}

\usepackage{paralist} % \asparaitem \compactitem \asparaenum \compactenum
\usepackage{verbatim} % for literal text input

\usepackage{fancybox}
\usepackage{varioref} % vref vpageref
\usepackage{chapterbib} % referece in each chapter
\usepackage{eurosym}
\usepackage[running]{lineno}
\usepackage{afterpage}

\usepackage[compatible]{nomencl}
\usepackage{xspace}
\newcommand{\dd}{\mathrm{d}} % for differential operator d
\newcommand{\abs}[1]{\left\lvert#1\right\rvert}
\newcommand{\norm}[1]{\left\lVert#1\right\rVert}

\newcommand{\pbk}[1]{\left(#1\right)}
\newcommand{\cbk}[1]{\left\lbrace#1\right\rbrace}
\newcommand{\sbk}[1]{\left\lbrack#1\right\rbrack}
\newcommand{\ie}{\textit{i.e.}\@\xspace}
\newcommand{\eg}{\textit{e.g.}\@\xspace}
\newcommand{\HE}{\hbox{H\kern-.12em\lower.48ex\hbox{E}}}
\def\leaderfill{\leaders\hbox to 1em{\hss.\hss}\hfill}
\DeclareSIUnit\molar{\mole\per\cubic\metre}
\DeclareSIUnit\Molar{\textsc{M}}

\graphicspath{ {images/} }
\begin{document}

    \title{Model-based process design of a ternary protein separation using multi-step gradient ion-exchange SMB chromatography}

    \author[1,2,3]{Qiao-Le He}
    \author[3]{Eric von Lieres}
    \author[4]{Zhaoxi Sun}
    \author[1,2]{Liming Zhao}
    \affil[1]{\textit{State Key Laboratory of Bioreactor Engineering, East China University of Science and Technology, 200237 Shanghai, China}}
    \affil[2]{\textit{R\&D Center of Separation and Extraction Technology in Fermentation Industry, East China University of Science and Technology, 200237 Shanghai, China}}
    \affil[3]{\textit{Forschungszentrum J\"ulich, IBG-1:Biotechnology, 52425 J\"ulich, Germany}}
    \affil[4]{\textit{Forschungszentrum J\"ulich, IAS-5/INM-9:Computational Biomedicine, 52425 J\"ulich, Germany}}
    \date{Apr.5, 2020}

    \maketitle

    \begin{abstract}
        Model-based process design of ion-exchange simulated moving bed (\textsc{iex-smb}) chromatography for center-cut separation of proteins is studied.
        Use of nonlinear binding models that describe more accurate adsorption behaviors of macro-molecules could make it impossible to utilize triangle theory to obtain operating parameters.
        Moreover, triangle theory provides no rules to design salt profiles in \textsc{iex-smb}.
        In the modeling study here, proteins (\ie, ribonuclease, cytochrome and lysozyme) on chromatographic columns packed with strong cation-exchanger SP Sepharose FF is used as an example system. 
        The general rate model with steric mass-action kinetics was used; two closed-loop \textsc{iex-smb} network schemes were investigated (\ie, cascade and eight-zone schemes).
        Performance of the \textsc{iex-smb} schemes was examined with respect to multi-objective indicators (\ie, purity and yield) and productivity, and compared to a single column batch system with the same amount of resin utilized.
        A multi-objective sampling algorithm, Markov Chain Monte Carlo (\textsc{mcmc}), was used to generate samples for constructing Pareto optimal fronts. 
        \textsc{mcmc} serves on the sampling purpose, which is interested in sampling the Pareto optimal points as well as those near Pareto optimal.
        Pareto fronts of the three schemes provide full information on the trade-off between the conflicting indicators, purity and yield here.
        The results indicate that the cascade \textsc{iex-smb} scheme and the integrated eight-zone \textsc{iex-smb} scheme have similar performance and that both outperforms the single column batch system.
    \end{abstract}

\section{Introduction}
\begin{figure}[h]
    \centering
    \includegraphics[width=0.5\textwidth]{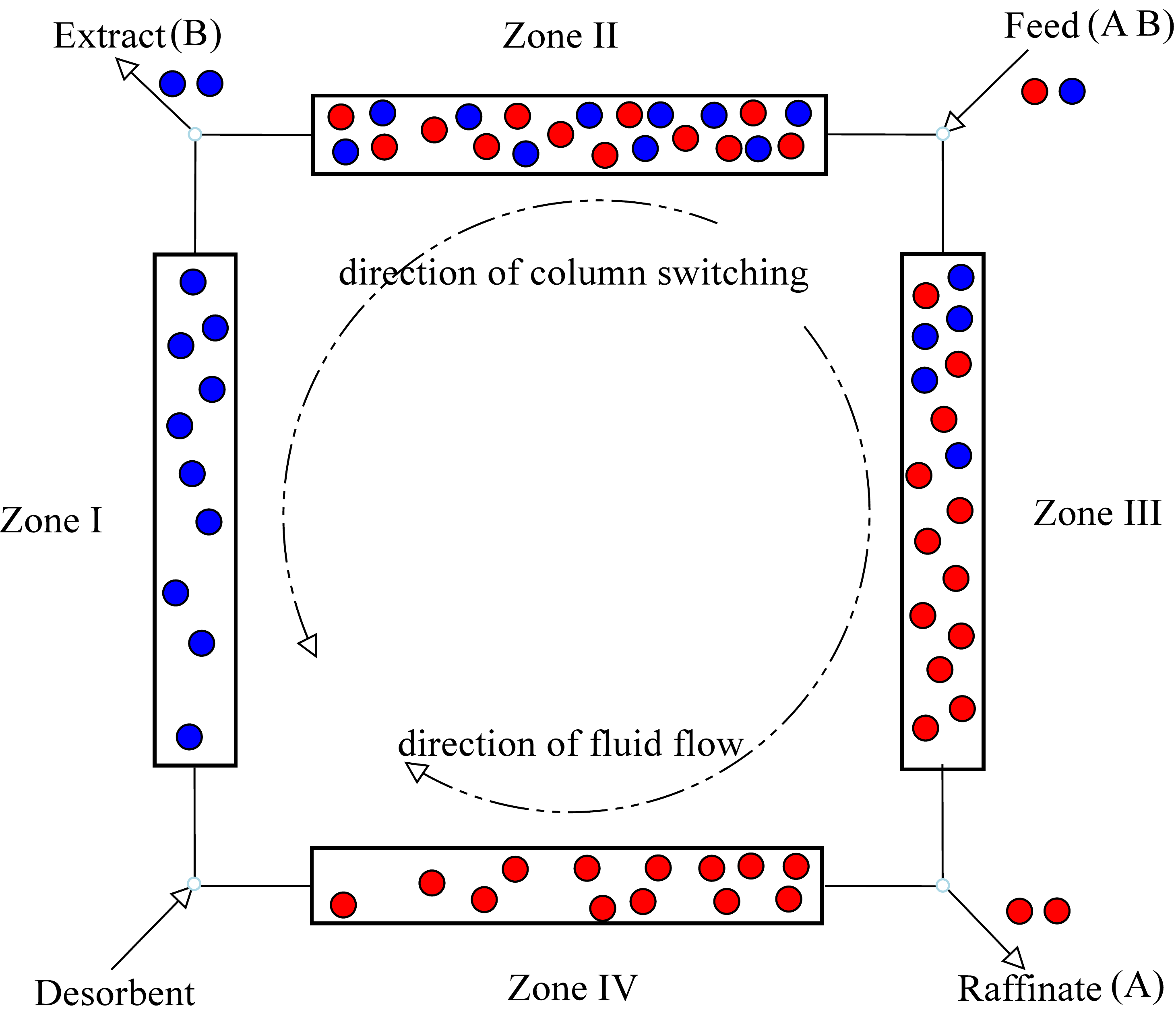}
    \caption{Schematic of four-zone \textsc{smb} chromatography}
    \label{fig:4-zone}
\end{figure}

Protein purification and separation are major concerns in the downstream processing of pharmaceutical industries \citep{carta2010protein,scopes2013protein}.
They require a series of processes aimed at isolating single or multiple types of proteins from complex intermediates of process productions. %, such as cells, tissues and organisms.
Choosing a purification method, depending on the purpose for which the protein is needed, is a critical issue.
Chromatography is a prevailing separation technology.
Simulated moving bed (\textsc{smb}) chromatography \citep{broughton1961continuous} is an excellent alternative to single column batch chromatography, because of its continuous counter-current operation, its potential to enhance productivities and to reduce solvent consumption \citep{seidel2008new, rajendran2009simulated, rodrigues2015simulated}.
According to the position of columns relative to the ports (\ie, feed, raffinate, desorbent, extract), the process is divided into four zones (\ie, I, II, III, IV), each has a specific functionality in the separations (see Fig.~\ref{fig:4-zone}).

Multicolumn continuous chromatography is also commonly used in protein separation and purification.
For instance, there are Japan Organo (\textsc{jo}) process (also termed as pseudo-\textsc{smb}) \citep{mata2001separation}, sequential multicolumn chromatography (\textsc{smcc}) \citep{ng2014design}, periodic counter-current packed bed chromatography (\textsc{pcc}) \citep{pollock2013optimising}, gradient with steady state recycle (\textsc{gssr}) process \citep{silva2010new}, multicolumn solvent gradient purification (\textsc{mcsgp}) \citep{aumann2007continuous} and capture \textsc{smb} \citep{angarita2015twin}.

However, the prominent features of \textsc{smb} processes are based on the prerequisite that optimal operating conditions can be determined, which can be very challenging in practice.
Determination of operating conditions is predominantly based on classical and extended triangle theories \citep{storti1993robust}, which are originally derived from the ideal chromatographic model with the linear isotherm.
Powerful criteria to design processes with nonlinear Langmuir pattern (\eg, Langmuir, modified Langmuir and bi-Langmuir isotherms) have been developed in the past years \citep{charton1995complete, mazzotti1997optimal, antos2001application}.
Versatile results on triangle theory for the design of \textsc{smb} processes have been published \citep{nowak2012theoretical, kim2016combined, lim2004optimization, kazi2012optimization, bentley2013prediction, bentley2014experimental, sreedhar2014simulated, toumi2007efficient, silva2015modeling, kiwala2016center}.

Substances separated by \textsc{smb} processes have recently been evolved from monosaccharides to proteins and other macro-molecules.
The adsorption behaviors of macro-molecules, as observed both in experiments \citep{clark2007new} and molecular dynamic simulations \citep{dismer2010structure, liang2012adsorption, lang2015comprehensive}, are much more complex than that described by the linear isotherm, the Langmuir pattern or the steric mass-action (\textsc{sma}) model \citep{brooks1992steric}.
Macro-molecules can undergo conformation changes in the mobile phase, orientation changes on the functional surfaces of the stationary phase and aggregation in the whole process.
\cite{clark2007new} used an acoustically actuated resonant membrane sensor to monitor the changes of surface energy and found that the changes persist long after mass loading of the protein has reached the steady state.
Therefore, more sophisticated adsorption models, such as the multi-state \textsc{sma} \citep{diedrich2017multi} based on the spreading model \citep{ghosh2013zonal, ghosh2014zonal}, have been proposed to describe dynamic protein-ligand interactions.
%After validating with experimental data, good agreements with experimental data and simulated data have been achieved.
Although these models describe more accurately the adsorption behaviors by taking orientation changes into consideration, the strong nonlinearity of the binding process causes tremendous difficulties in deriving analytical formulas to calculate flowrates, as required in the triangle theory.
%Hence, triangle theory can not be applied to \textsc{smb} processes.
%to determine operating conditions of \textsc{smb} processes to separate macro-molecules.

Ion-exchange (\textsc{iex}) mechanism is commonly used to separate charged components in chromatography.
In \textsc{iex} chromatography (both single column and \textsc{smb} processes), an auxiliary component acting as a modifier is used to change electrostatic interaction forces between functional groups and macro-molecules; sodium chloride is most frequently used.
The higher the salt concentration is, the lower the electrostatic force is, such that the binding affinity is decreased and bound components are successively eluted \citep{lang2015comprehensive}.
In the single column \textsc{iex} processes, this principle can be implemented in an isocratic mode or in a linear gradient mode.
The linear gradient mode has frequently been applied to achieve better separation capability \citep{osberghaus2012optimizing}.
While in the \textsc{iex-smb} processes, the isocratic mode (\ie, the salt strength is identical in all zones) is originally applied.
The gradient mode can be applied, by using a salt strength in the desorbent stream that is stronger than that in the feed (cf.~Fig.~\ref{fig:4-zone}), thus, high salt concentration in zones I and II (high salt region), low salt concentration in zones III and IV (low salt region).
%The difference of salt profiles in the two distinct regions of \textsc{iex-smb} are defined as salt gradient.
%It was then realized beneficial to have gradient mode in \textsc{iex-smb}, by implementing a salt strength in the desorbent stream that is stronger than that in the feed.
As shown in some studies, applying the gradient mode can further improve the separation performance \citep{antos2001application, houwing2002effect, houwing2003positioning, li2007proteins}.
The salt profiles of the gradient mode are referred to as two-step salt gradient in this work; an extension to multi-step gradients for ternary separations is straightforward.
However, the construction of the multi-step salt gradient imposes difficulty on the design of \textsc{iex-smb} processes.

This difficulty can be detoured by using open-loop multicolumn chromatography (\eg, \textsc{jo} and \textsc{mcsgp}), where it is possible to apply linear salt gradients.
The open-loop feature and possibility of linear gradient implementation provides great application potential for multicolumn continuous chromatography in downstream processing \citep{faria2015instrumental}, as dealing with the model complexity induced by linear gradients is much easier than that by multi-step ones; moreover, the open-loop feature renders more degrees of freedom in the process designs.

Conventional four-zone \textsc{smb} processes are tailored for binary separations.
Ternary separations are currently requested in many applications; the selection of network configurations is application-specific.
Ternary separations, also referred to as center-cut separations, can be performed  by the \textsc{jo} process which operates in semi-continuous mode exploiting batch chromatography followed by the \textsc{smb} operation without feed delivery \citep{borges2008design, gracca2015separation}. 
They can also be achieved by cascading two four-zone \textsc{smb} units \citep{wooley1998nine, nicolaos2001application, nowak2012theoretical}.
Cascade schemes require designing and operating two four-zone \textsc{smb} processes, which can be laborious and costly.
Hence, integrated schemes have been developed using a single multi-port valve and fewer pumps than cascade schemes \citep{seidel2008new, da2016evaluation}.
Moreover, process design is simplified due to simultaneous switching of all columns.
%In network configurations for ternary separations, multi-step salt gradients rather than two-step ones have to be designed, where a comprehensive column model, fast numerical solver and efficient optimizer are desired.
As the complexity of network configurations, in addition to the nonlinearity of the binding models and multi-step gradient, a comprehensive column model, fast numerical solver and efficient optimizer are required for the process design.

In this study, we shall present a model-based process design of \textsc{iex-smb} units for separating a protein mixture of ribonuclease, cytochrome and lysozyme, using cation-exchange columns packed with SP Sepharose FF beads.
Two network configurations for the ternary separation will be used, that is, a cascade scheme and an eight-zone scheme.
For comparison, a conventional single column batch system, with the same column geometry and resin utilization, will be studied.
In this study, \textsc{smb} processes are modeled by weakly coupling individual models of the involved columns \citep{he2018efficient}.
%The code has been published as open-source software, \textsc{cadet-smb}.

The operating conditions of the network configurations will be optimized by a stochastic algorithm, Markov Chain Monte Carlo (\textsc{mcmc}), with respect to conflicting objectives of purity and yield.
Pareto fronts will be computed for illustrating the best compromises between the two conflicting performance indicators.
Unlike multi-objective optimization algorithms (\eg, non-dominated sorted genetic algorithm, strength Pareto evolutionary algorithms) that try to eliminate all the non-dominated points during optimization, \textsc{mcmc} serves on the sampling purpose, which is interested in sampling the Pareto optimal points as well as those near Pareto optimal.
For sampling purpose, \textsc{mcmc} not only accepts proposals with better objective values, but also accepts moves heading to non-dominated points with certain probability.
Therefore, \textsc{mcmc} could render greater information for process design, such as, uncertainties of parameters that is related to the robustness of process design can be examined. % \citep{he2019bayesian}. %, though the information is not involved in this study.
As shown in the Pareto fronts, the performance indicators of the single column are dominated by that of both the cascade and the eight-zone schemes.
The performances of the two \textsc{iex-smb} schemes are quite similar, though the cascade scheme has a slight advantage over the eight-zone scheme.

\section{Theory}
This section introduces the transport model, the binding model, the load-wash-elution mode of single column chromatography, the \textsc{smb} network connectivity, the performance indicators and the optimization algorithm.

\subsection{Transport model}
The transport of proteins in the column is described by means of the general rate model (\textsc{grm}), which accounts for various levels of mass transfer resistance \citep{guiochon2006fundamentals}.

\textsc{grm} considers convection and axial dispersion in the bulk liquid, as well as film mass transfer and pore diffusion in the porous beads:
\begin{subequations}\label{eq:GRM}
    \begin{align}
        \frac{\partial c_i^j}{\partial t} & = -u_{\text{int}}^j \frac{\partial c_i^j}{\partial z} + D_{\text{ax}}^j \frac{\partial^2 c_i^j}{\partial z^2} - \frac{1-\varepsilon_c}{\varepsilon_c} \frac{3}{r_p} k_{f,i}^j \left(c_i^j - c_{p,i}^j(r\!=\!r_p)\right) \\
    \frac{\partial c_{p,i}^j}{\partial t} & = D_{p,i}^j \left(\frac{\partial^2 c_{p,i}^j}{\partial r^2} + \frac{2}{r}\frac{\partial c_{p,i}^j}{\partial r}\right) - \frac{1-\varepsilon_p}{\varepsilon_p}\frac{\partial q_i^j}{\partial t}
    \end{align}
\end{subequations}
where $c_i^j$, $c_{p,i}^j$ and $q_i^j$ denote the interstitial, stagnant and stationary phase concentrations of component $i\in \{1,\dots,M\}$ in column $j \in \{1, \dots, N\}$.
Particularly in the \textsc{iex} chromatography, when the salt component is introduced, $i\in\{0, 1, \dots, M\}$ and $i=0$ denotes the salt ions, whose concentration will be required by the steric mass-action (\textsc{sma}) model (cf.~Eq.~\eqref{eq:sma}).
Furthermore, $z \in [0,L]$ denotes the axial position where $L$ is the column length, $r \in [0,r_p]$ is the radial position, $r_p$ is the particle radius, $t$ time, $\varepsilon_c$ column porosity, $\varepsilon_p$ particle porosity, $u_{\text{int}}^j$ interstitial velocity, $D_{\text{ax}}^j$ is the axial dispersion coefficient, $D_{p,i}^j$ the effective pore diffusion coefficient, and $k_{f,i}^j$ the film mass transfer coefficient.
At the column inlet and outlet, Danckwerts boundary conditions \citep{Barber1998Boundary} are applied:
\begin{equation}
    \begin{cases}
        \left. \dfrac{\partial c_i^j}{\partial z} \right|_{z=0} &= \dfrac{u_{\text{int}}^j}{D_{\text{ax}}^j} \left(c_i^j(z\!=\!0) - c_{\text{in},i}^j \right) \\
        \left. \dfrac{\partial c_i^j}{\partial z} \right|_{z=L} &= 0
    \end{cases}
    \label{eq:Danckwerts_column}
\end{equation}
where $c_{\text{in},i}^j$ is the inlet concentration of component $i$ in column $j$; the calculation of inlet concentration in \textsc{smb} chromatography is deferred to Eq.~\eqref{eq:node_balance}.
The boundary conditions at the particle surface and center are given by:

\begin{equation}
    \begin{cases}
        \left. \dfrac{\partial c_{p,i}^j}{\partial r} \right|_{r=r_p} &= \dfrac{k_{f,i}^j}{\varepsilon_p D_{p,i}^j} \left(c_i^j - c_{p,i}^j (r\!=\!r_p)\right) \\
        \left. \dfrac{\partial c_{p,i}^j}{\partial r} \right|_{r= 0} &= 0
    \end{cases}
    \label{eq:Danckwerts_surface}
\end{equation}

\subsection{Binding model}
%Due to the complexity of protein-ligand interaction, the knowledge of adsorption kinetics is of the most importance. 
The \textsc{sma} model has widely been used for predicting and describing nonlinear ion-exchange adsorption of proteins.
The binding of components $i \geqslant 1$ is expressed as $\frac{\partial q_i^j}{\partial t} = f(c_{p,i}^j, q_i^j, c_{p,0}^j, q_0^j)$, \ie, the adsorptive fluxes of components $i \geqslant 1$ are influenced by the salt concentrations, $q_0^j, c_{p,0}^j\, (i = 0)$:
%; the relationship between the stagnant liquid $c_{p,i}^j$ and the stationary phase $q_i^j$ is described as follows:
\begin{equation}
\frac{\partial q_i^j}{\partial t} = k_{a,i}\, c_{p,i}^j\, (\bar{q}_0^j)^{\nu_i} - k_{d,i}\, q_i^j\, (c_{p,0}^j)^{\nu_i}
    \label{eq:sma}
\end{equation}
where $k_{a,i}$ and $k_{d,i}$ denote adsorption and desorption coefficients, and $\nu_i$ is the characteristic charge of the adsorbing molecules.
%$c_0$, $c_{p,0}$, $q_0$ denote the salt concentration in the three different phases.
The concentration $\bar{q}_0^j$ of available counter ions has to be algebraically determined from an electro-neutrality condition, as Eq.~\eqref{eq:sma} is only valid for components $i \geqslant 1$, 
%denotes counter ions that are not shielded and available for protein binding,
\begin{equation}
    \bar{q}_0^j = \Lambda - \sum_{i=1}^{M} (\nu_i + \sigma_i) q_i^j
\end{equation}
Here $\Lambda$ is the ionic capacity, and $\sigma_i$ the shielding factor.
Then, the salt concentration in the stationary phase is given by:
\begin{equation}
    q_0^j = \bar{q}_0^j + \sum_{i=1}^M \sigma_i q_i^j
\end{equation}

\subsection{Single column batch system}
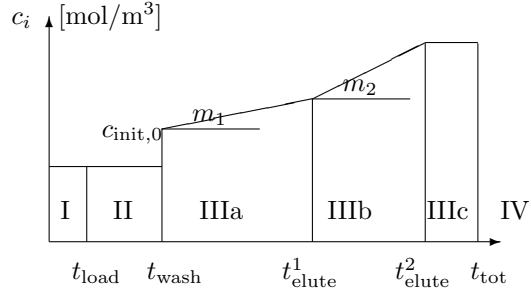
\begin{figure}[h]
    \centering
    \begin{minipage}[b]{0.5\textwidth}
        \setlength{\unitlength}{1cm}
        \begin{picture}(6,4)
            \put(0.5,0.5){\vector(0,1){3}}
            \put(0.5,0.5){\vector(1,0){6}}
            \put(0,3.4){$c_i \quad [\si{\mole\per\cubic\metre]}$}
            \put(6.1,0){$t_{\text{tot}}$}

            \put(0.5,1.5){\line(1,0){1.5}}
            %\put(0,1.4){\num{50}}
            \put(1,0.5){\line(0,1){1}}
            \put(0.8,0){$t_{\text{load}}$}
            \put(0.65,0.8){I}

            \put(2,0.5){\line(0,1){1.5}}
            \put(1.8,0){$t_{\text{wash}}$}
            \put(1.35,0.8){II}
            \put(1.2,1.9){$c_{\text{init},0}$}

            \put(2,2){\line(5,1){2}}
            \put(2,2){\line(1,0){1.3}}
            \put(2.4,2.1){$m_1$}

            \put(4,0.5){\line(0,1){1.9}}
            \put(4,2.4){\line(1,0){1.3}}
            \put(4,2.4){\line(2,1){1.5}}
            \put(4.4,2.5){$m_2$}
            \put(3.6,0){$t_{\text{elute}}^1$}
            \put(2.5,0.8){IIIa}
            \put(4.2,0.8){IIIb}
            \put(5.52,0.8){IIIc}
            \put(6.5,0.8){IV}

            \put(5.5,0.5){\line(0,1){2.64}}
            \put(5.5,3.15){\line(1,0){0.7}}
            \put(5.1,0){$t_{\text{elute}}^2$}
            \put(6.2,0.5){\line(0,1){2.64}}
        \end{picture}
    \end{minipage}
    \caption{Sequential phases of single column gradient elution chromatography. I: load phase; II: wash phase; III: multi-step elution phase; IV: regeneration.}
    \label{fig:batch_salt_scheme}
\end{figure}

There are various possibilities to operate a single column in batch manner.
The specific one described here is used for comparison with the \textsc{smb} processes; we refer you the protocol to \cite{osberghaus2012optimizing}.
Apart from regeneration steps, an operating cycle of this single column process consist of three phases: load, wash and elution, see Fig.~\ref{fig:batch_salt_scheme}.
In phase I, the column is equilibrated with the running buffer, then the protein solution is injected to the column for time $t_{\text{load}}$.
This is followed by a wash step (II) for time $t_{\text{wash}}$.
Afterwards, different gradients including a gradient of zero length, \ie, isocratic elution, are set to elute the bound protein for time $t_{\text{elute}}$.
The elution phase (III) can be implemented by multi-step solvent gradients.
As depicted in Fig.~\ref{fig:batch_salt_scheme} as an example, there are two parts (\ie, IIIa, IIIb) with linear gradients but different slopes ($m_1$, $m_2$), and eventually an isocratic part (IIIc).
Multi-step gradients are used, because in center-cut separations peak overlaps at the front and at the back of the target peak need to be minimized.
The regeneration of the column (IV) completes one operating cycle of the single column batch system.

\subsection{SMB network connectivity}
In \textsc{smb} chromatography two adjacent columns, $j$ and $j\!+\!1$, are connected via a node $j$, see Fig.~\ref{fig:adjacent}.
A circular \textsc{smb} loop is closed when two column indices point to the same physical column (\ie, by identifying column $j=N\!+\!1$ with column $j=1$), see Fig.~\ref{fig:4-zone}.
In this work, node $j$ is located at the downstream side of column $j$ and the upstream side of column $j\!+\!1$.
At each node only one or none of the feed (F), desorbent (D), raffinate (R), or extract (E) streams exists at a time.
The inlet concentration of component $i \in \{0, 1, \dots, M\}$ in column $j\!+\!1$ is calculated from mass balance of the node $j$:

\begin{figure}
    \centering
    \includegraphics[width=0.5\textwidth]{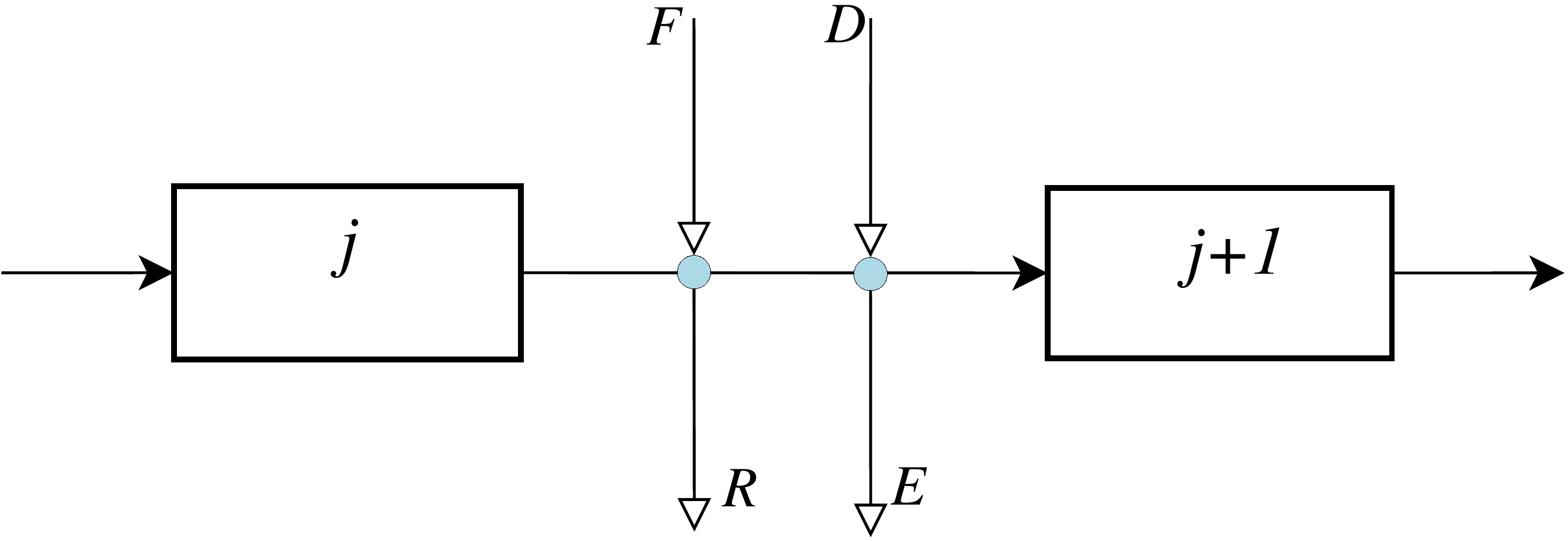}
    \caption{Schematic of network connection between two adjacent columns.}
    \label{fig:adjacent}
\end{figure}

\begin{align}
    c_{\text{in},i}^{j+1} = \frac{c_{\text{out},i}^j Q^j + \delta_i^j}{Q^{j+1}}
    \label{eq:node_balance}
\end{align}
where $c_{\text{out},i}^j = c_i^j(t, z\!=\!L)$ denotes the outlet concentration of component $i$ in column $j$, $Q^j = \varepsilon_c u_{\text{int}}^j \pi d_c^2/4$ the volumetric flowrates and $d_c$ the column diameter.
Meanwhile, $\delta^j$ determines the current role of node $j$ (\ie, F, D, R, E or none):

\begin{align}
    \delta^j_i = \left\{  \begin{array}{l@{\quad \quad}l}
            \phantom{-}c_{\text{in},i}^F\, Q^F & \text{feed} \\
            \phantom{-}c_{\text{in},i}^D\, Q^D & \text{desorbent} \\
            -c_{\text{out},i}^j\, Q^R & \text{raffinate} \\
            -c_{\text{out},i}^j\, Q^E & \text{extract} \\
        \phantom{-}0       & \text{none}
    \end{array} \right.
    \label{eq:delta}
\end{align}
where $c_{\text{in}}^F$ and $c_{\text{in}}^D$ are the component concentrations at feed and desorbent ports, and $Q^F$, $Q^D$, $Q^R$, $Q^E$ the volumetric flowrates at the feed, desorbent, raffinate, and extract ports, respectively.
$\delta^j = 0$ indicates nodes that are currently not connected to a port (\ie, in the interior of a zone); this occurs when more columns than zones are present, such as eight columns in a four-zone scheme.
As there might be multiple feed, desorbent, raffinate, and extract ports in \textsc{smb} processes (\eg, the cascade scheme and the integrated eight-zone scheme), an straightforward extension of the indexing scheme is required.
Column shifting is implemented by periodically permuting $\delta$ each switching time $t_s$.

\subsection{Performance indicators}
Purity, yield and productivity are commonly used for evaluating the performance of chromatographic processes.
All indicators can be defined within one collection time $t_c$ \citep{bochenek2013evaluating}.
In the case of \textsc{smb} systems, $t_c$ is the switching time $t_s$, while in single column systems it is the length of pooling time interval $t_p$.

In this study, performance indicators are all defined in terms of component, $i \in \{1, \dots, M\}$, withdrawn at a \textsc{smb} node, $j \in \{E, R, \dots\}$, or collected at the outlet of single column systems, $j \in \{B\}$, within the pooling interval.
Note that we do not calculate the performance indicators of the salt component, $i = 0$.
The definitions of performance indicators are all based on concentration integrals $A_{\text{out},i}^j$ of component $i$ at node $j$,
\begin{equation}
    A_{\text{out},i}^j = \int_{t=\tau}^{\tau + t_c} c_i^j(t, z\!=\!L)\, \dd t
\end{equation}
In \textsc{smb} systems, $\tau$ is the starting time of one switching interval and all performance indicators are calculated when the system is upon cyclic steady state (\textsc{css}).
In single column systems, $\tau$ is the starting point of the pooling time interval.

The purity, $\mathtt{Pu}_{i}^j$, is the concentration integral of component $i$ relative to the integral sum of all components, Eq.~\eqref{eq:purity}. 
The yield, $\mathtt{Y}_i^j$, is the ratio of the withdrawn mass and the feed mass, Eq.~\eqref{eq:yield}.
\begin{equation}
    \mathtt{Pu}_{i}^j = \frac{A_{\text{out},i}^j }{\sum\limits_{k=1}^{M} A_{\text{out},k}^j}
    \label{eq:purity}
\end{equation}
\begin{equation}
    \mathtt{Y}_i^j = \frac{Q^j A_{\text{out},i}^j}{Q^F c_{\text{in},i}^F\, t_{\text{load}}}
    \label{eq:yield}
\end{equation}
In \textsc{smb} systems, when the solution stream is continuously fed via the feed node and the extracts and raffinates are continuously withdrawn, $t_{\text{load}} = t_s = t_c$ and $Q^F$ differs from $Q^j$.
While in single column systems, the flowrates are the same at the inlet node and outlet node, $Q^j = Q^F$; and $t_{\text{load}} \neq t_p = t_c$.
The productivity, $\mathtt{Pr}_{i}^{j}$, is the withdrawn mass of the component $i$ per collection time relative to the total volume of the utilized packed bed in all columns,
\begin{equation}
    \mathtt{Pr}_{i}^{j} = \frac{Q^j A_{\text{out},i}^j}{t_c\, (1-\varepsilon_c) V_c\, N}
\end{equation}
In single column systems, $N = 1$.
According to the definitions of $\mathtt{Pr}_i^j$ and $\mathtt{Y}_i^j$, both indicators increase with increasing amounts of the product collected, \ie, $A_{\text{out},i}^j$.
However, $\mathtt{Pr}_i^j$ can also be improved by reducing the collection time $t_c$.

\subsection{Optimization}
Generally, operating conditions of both \textsc{smb} and single column processes are systematically optimized by numerical algorithms such that the above performance indicators are all maximized.
In studies of center-cut separations, the middle component $i$ is of interest. 
To specifically account for a vector of the conflicting performance indicators of $\sbk{\mathtt{Pu}_i^j, \mathtt{Y}_i^j}$ (referred to as $\mathcal{PI}$ hereafter) in this study, multi-objective optimization is applied.

A set of objectives can be combined into a single objective by optimizing a weighted sum of the objectives \citep{marler2010weighted}, or keeping just one of the objectives and with the rest of the objectives constrained (the $\varepsilon$-constraint method \citep{mavrotas2009effective}).
The latter method is used in this work, that is, maximizing the yield of the target component of the processes with the purity constrained to be larger than a threshold $\varepsilon_i^j$:
\begin{equation}
    \begin{array}[ ]{r l}
        \min        & f(\theta) = - \mathtt{Y}_i^{j},\quad\, i\in\{1, \dots, M\} \\
        \text{s.t.} & \begin{cases}
            c_i^j(\theta) : \mathtt{Pu}^j_i - \varepsilon_i^j \geqslant 0 \\
            \theta_{\min} \leqslant \theta \leqslant \theta_{\max}
        \end{cases}
    \end{array}
    \label{eq:eps_constraint}
\end{equation}
where $\theta$ is a vector of optimized parameters with boundary limitations of $\sbk{\theta_{\min}, \theta_{\max}}$.
The inequality $c_i^j(\theta)$ is lumped into the objective function using penalty terms in this study, such that it can be solved as a series of unconstrained minimization problems with increasing penalty factors, $\sigma_k$,
\begin{equation}
    \min \mathcal{H}(\theta; \sigma_k) = f(\theta) + \sigma_k g(\theta)
    \label{eq:objective}
\end{equation}
In Eq.~\eqref{eq:objective}, the penalty function is chosen as $g(\theta) = \sum_{j\in\{R,E\}} \norm{ \min\{0, c^j(\theta)\} }^2$.
%Note that $\mathcal{H}(\theta;\sigma_k)$ is not differentiable at some $\theta$ because the chosen quadratic penalty term is not continuous.
%Hence, the optimization problem here is a quadratic non-smooth penalty function.

Different types of methods can be applied to solve the minimization of $\mathcal{H}(\theta;\sigma_k)$, such as deterministic methods and heuristic methods. 
Additionally, stochastic methods can be chosen.
In order to use a stochastic method, the minimization of $\mathcal{H}(\theta;\sigma_k)$ is further formulated to maximum likelihood estimation:
\begin{equation}
    \text{arg}\max_{\theta}\, \mathcal{L}(\theta)
    \label{eq:mle}
\end{equation}
where the likelihood is defined as an exponential function of $\mathcal{H}(\theta;\sigma_k)$:
\begin{equation}
    \mathcal{L}(\theta) \overset{\text{def}}= \exp\sbk{ - \frac{1}{2} \mathcal{H}(\theta; \sigma_k) }
    \label{eq:likelihood}
\end{equation}

\subsection{Numerical solution}
All numerical simulations were computed on an Intel(R) Xeon(R) system with 16 CPU cores (64 threads) running at \SI{2.10}{\giga\hertz}.

The mathematical models described above for each column of \textsc{smb} processes are weakly coupled together and then iteratively solved.
The open-source code has been published, \url{https://github.com/modsim/CADET-SMB}.
\textsc{cadet-smb} repeatedly invokes \textsc{cadet} kernel to solve each individual column model, with default parameter settings.
\textsc{cadet} is also an open-source software, \url{https://github.com/modsim/CADET}.
The axial column dimension is discretized into $N_z = 40$ cells, while the radical bead is discretized into $N_r = 10$ cells.
The resulting system of ordinary differential equations is solved using an absolute tolerance of $\num{1e-10}$, relative tolerance of $\num{1e-6}$, an initial step size of $\num{1e-14}$ and a maximal step size of $\num{5e6}$.

%Differential evolution (\textsc{de}) \citep{storn1997differential} algorithm was employed in the optimization design because of its simple and fast features.
A stochastic multi-objective sampling algorithm, \textsc{mcmc}, is applied in this study to optimize the operating conditions.
A software computing the Metropolis-Hastings algorithm with delayed rejection, adjusted Metropolis and Gibbs sampling has been published as open-source software, \url{https://github.com/modsim/CADET-MCMC}.
For sampling purpose, \textsc{mcmc} not only accepts proposals with better objective values, but also accepts moves heading to non-dominated points with certain probability.
%To be specific, new proposed samples are either accepted or rejected according to the likelihood-odds, $\gamma = \frac{\mathcal{L}(\hat{\theta})}{\mathcal{L}(\theta)}$. 
%Thereafter, performance indicators of the samples are calculated via $\mathcal{PI} = f(\Theta)$.
%In the multi-objective optimization, Pareto samples can be further accumulated by repeatedly solving Eq.~\eqref{eq:mle} with varying values of $\varepsilon_i^j$.
Samples are collected until the Geweke convergence criterion \citep{geweke1991evaluating} is smaller than $\num{1e-4}$ or the desired sample size is reached.
%The Pareto fronts in this study describe two-dimensional trade-offs between purity and yield.
Non-dominated stable sort method of Pareto fronts were is applied to generate the frontiers \citep{duh2012learning}.

\section{Case}
A protein mixture of ribonuclease, cytochrome and lysozyme is of interest in this study, $i\in \{\text{RNase, cyt, lyz}\}$.
Sodium chloride, $i = 0$, acts as the modifier of binding affinities of the proteins.
It is a prototype example used in the academic field for modeling purposes \citep{osberghaus2012determination, osberghaus2012optimizing}.
%The fractionation of this protein mixture is also referred to as center-cut separation, that is, 
The center component, cyt, is targeted and the other two components are regarded as impurities.
We have modeled separation of these proteins on chromatographic columns packed with strong cation-exchanger SP Sepharose FF (see the column geometry in Tab.~\ref{tab:literature_data}).
%The running buffer is $\SI{20}{\milli\molar}$ sodium phosphate buffer at pH 7.

\begin{table}[h]
    \centering
    \scriptsize
    \caption{Parameters of column geometry, mass transport and binding of the model proteins ($i \in \{\text{RNase, cyt, lyz}\}$).}
    \label{tab:literature_data}
    \begin{tabular}{c c l l S}
        \toprule
        Catalog                     & Symbol              & Description               & {Value}             & {Unit} \\
        \midrule
        \multirow{5}{*}{Geometry}   & $L$                 & column length             & $1.4\times10^{-2}$  & \si{\metre} \\
                                    & $d_c$               & column diameter           & $1\times10^{-2}$    & \si{\metre} \\
                                    & $d_p$               & particle diameter         & $9.00\times10^{-5}$ & \si{\metre} \\
                                    & $\varepsilon_b$     & column porosity           & 0.37                & \\
                                    & $\varepsilon_p$     & particle porosity         & 0.75                & \\

        \midrule
        \multirow{3}{*}{Transport}  & $D_{\text{ax}}$     & axial dispersion          & $5.75\times10^{-8}$ & \si{\square\metre\per\second} \\
                                    & $D_{p,i}$             & pore diffusion            & $6.07\times10^{-11}$ & \si{\square\metre\per\second} \\
                                    & $k_{f,i}$               & film mass transfer        & $6.90\times10^{-6}$ & \si{\metre\per\second} \\

        \midrule
        \multirow{5}{*}{Binding}    & $\Lambda$           & ionic capacity            & 1200                & \si{\mole\per\cubic\metre} \\
                                    & $k_{a,i}$           & adsorption coefficients   & $\sbk{7.70,\, 1.59,\, 35.5}$ & \si{\per\second}\\
                                    & $k_{d,i}$           & desorption coefficients   & $\sbk{1000,\, 1000,\, 1000}$ & \si{\per\second} \\
                                    & $\nu_i$             & characteristic charges    & $\sbk{3.70,\, 5.29,\, 4.70}$  & \\
                                    & $\sigma_i$          & steric factors            & $\sbk{10.0,\, 10.6,\, 11.83}$ & \\
        \bottomrule
    \end{tabular}
\end{table}
 
Simplified models (\eg, \textsc{tdm} and \textsc{edm}) that assume instance equilibrium of the binding process can not be applied here,
because mass transfer dynamics of macro-molecules in ca.~\SI{100}{\micro\metre} beads can be rate limiting \citep{lodi2017ion}.
Hence, the \textsc{grm} model is used to describe the mass transfer in the porous beads.
In addition, the \textsc{sma} model inherently considers the impact of salt on binding affinity, the hindrance effect of macro-molecules.
The mathematical modeling of the prototype example has been presented in \cite{montesinos2005analysis, puttmann2013fast, osberghaus2012determination, osberghaus2012optimizing}; the transport and binding parameters from experiments are shown in Tab.~\ref{tab:literature_data}.
The transport parameters can also be calculated from correlation formulas.
$D_\text{ax} = \SI{5.75e-8}{\square\metre\per\second}$, $D_p = \SI{6.07e-11}{\square\metre\per\second}$ and $k_f = \SI{6.90e-6}{\metre\per\second}$ are taken for all proteins because they have almost the same molecular weight.
Relatively large values are used for the rate parameters of the binding model, $k_{a,i}$ and $k_{d,i}$, as the adsorption-desorption kinetics is typically much faster than kinetics of mass transport in pores.
The geometry, transport and binding parameters are directly used in this study, while the operating conditions will be optimized.

In pharmaceutical manufacturing typically at least $99\%$ purity of cyt is needed.
Therefore, taking expensive computation cost in \textsc{iex-smb} simulations to generate Pareto optimal points with lower purities is not required here.
In this study, the $\varepsilon_{\text{cyt}}^j$ in Eq.~\eqref{eq:eps_constraint} is set to $99\%$; thus, only the Pareto optimal fragment with cyt purity higher than $99\%$ is concerned in \textsc{iex-smb}.
This implementation reduces the heavy computational burden of multi-objective sampling, as much fewer samples are requested to construct the frontiers.
Pareto optimal fragment can be further extended by repeatedly solving Eq.~\eqref{eq:mle} with varying values of $\varepsilon_{\text{cyt}}^j$.
This can be one of the advantages of the $\varepsilon$-constraint method.
In single column systems, the whole Pareto optimal front is concerned as the computational cost is not expensive.

\subsection{Single column}
In the single column system, the equilibration, load-wash-elution and regeneration phases are performed.
The interstitial velocity is \SI{5.75e-4}{\metre\per\second}, resulting in a retention time for a non-retained component of \SI{24.35}{\second}.
A sample with \SI{1}{\mole\per\cubic\metre} concentration of RNase, cyt and lyz is injected in the load phase.
In both the load and wash steps, a salt concentration of \SI{50}{\mole\per\cubic\metre} is applied. 
The operating time intervals for the load and wash phases are \SI{10}{\second} and \SI{40}{\second}.
The shapes of the multi-step elution phase, that can be characterized by the bilinear gradients (\ie, $m_1$ and $m_2$), the operating time intervals, and the initial salt concentration (\ie, $c_{\text{init},0}$), are optimized.
Fig.~\ref{fig:batch_salt_scheme} shows the elapsed times, from which the operating time intervals are calculated (\ie, $\Delta t_1 = t_{\text{elute}}^1 - 50$ and $\Delta t_2 = t_{\text{elute}}^2 - t_{\text{elute}}^1$).
Thus, the optimized operating parameters are $\theta = \{\Delta t_1, \Delta t_2, m_1, m_2, c_{\text{init},0}\}$.

\subsection{SMB}
Depending on initial states, \textsc{smb} processes often undergoes a ramp-up phase and eventually enter into a \textsc{css}.
In this study, the system is assumed to be upon \textsc{css} when a difference between two iterations $(k, k\!+\!1)$ falls below a predefined tolerance error, $e_t$, see Eq.~\eqref{eq:difference_method}.
%Upon \textsc{css}, performance indicators of \textsc{iex-smb} processes are calculated.
\begin{equation}
    \max_{j \in \{R,E\}} \sum_{i=1}^M \left( \int_0^{t_s} \left| c_{i,k}^j(t, z=L) - c_{i,k+1}^j(t, z=L) \right|^n \,\mathrm{d}t \right)^\frac{1}{n} \leqslant e_t 
    \label{eq:difference_method}
\end{equation}
In this study, the columns of \textsc{iex-smb} are initially empty except for the concentration of bound salt ions $q_0^j$ that is set to the ionic capacity $\Lambda$, in order to satisfy the electro-neutrality condition.
Additionally, the initial salt concentrations in column $j$, $c_0^j(t\!=\!0, z)$, $c_{p,0}^j(t\!=\!0, z)$, $q_0^j(t\!=\!0, z)$, are set equal to the salt concentrations in the respective upstream inlet nodes.
Taking the four-zone scheme as an example (cf.~Fig.~\ref{fig:4-zone}), the salt concentrations of the mobile and stationary phases ($c_0^j$ and $c_{p,0}^j$) in zone III and IV are set to the value of $c_0^F$; while $c_0^j$ and $c_{p,0}^j$ in zone I and II are set to the value of $c_0^D$:
\begin{equation}
    \begin{cases}
        c_0^j(t\!=\!0, z) = c_{p,0}^j(t\!=\!0, z) = c_0^D \qquad j \in \{\text{I}, \text{II}\} \\
        c_0^j(t\!=\!0, z) = c_{p,0}^j(t\!=\!0, z) = c_0^F \qquad j \in \{\text{III}, \text{IV}\} \\
        q_0^j(t\!=\!0, z) = \Lambda \qquad j \in\{\text{I}, \text{II}, \text{III}, \text{IV}\} 
    \end{cases}
    \label{eq:salt_initial}
\end{equation}
In contrast to the single column system, the chosen initial state of \textsc{smb} processes is irrelevant to the performance indicators calculated upon \textsc{css}.

\begin{figure}
    \centering
    \begin{subfigure}{\textwidth}{\includegraphics[width=\textwidth]{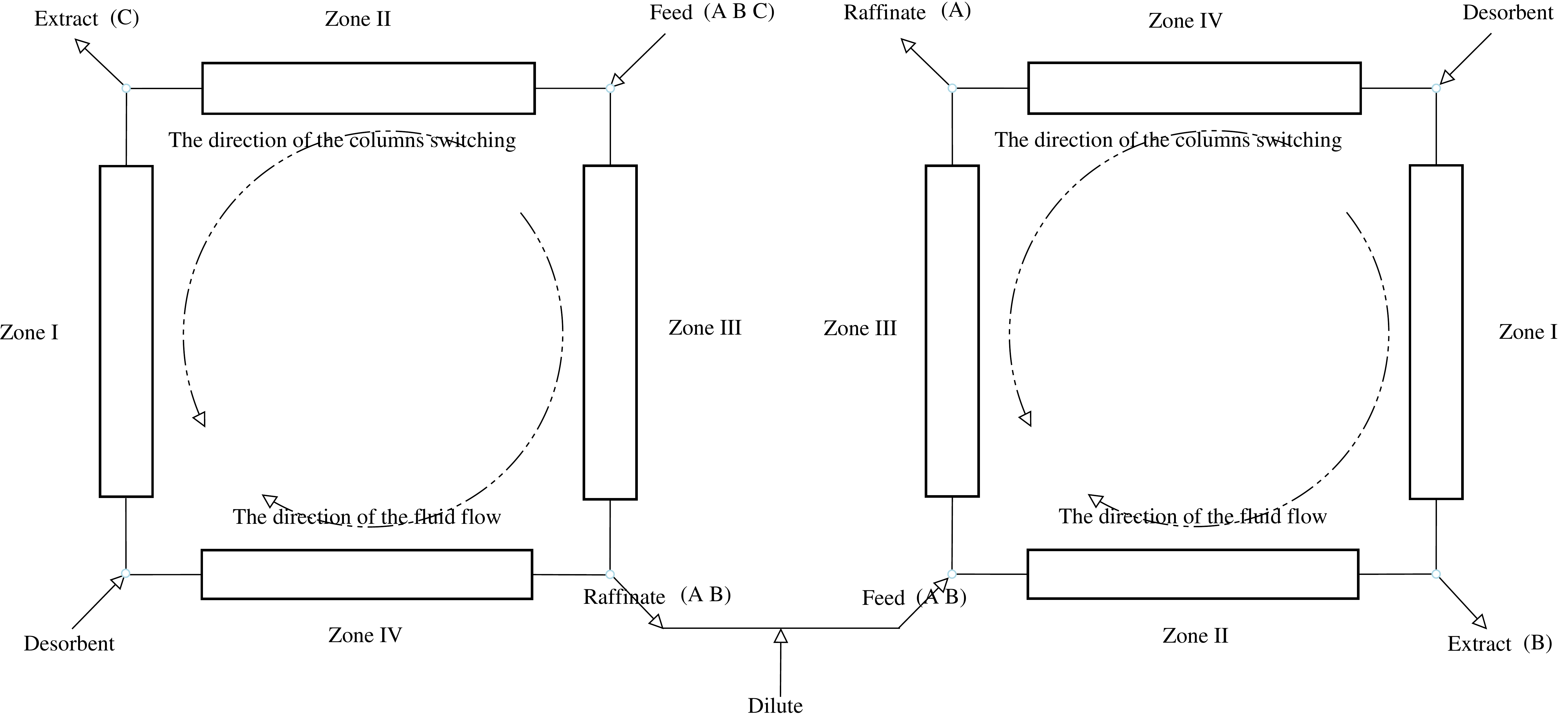}} \caption{} \label{} \end{subfigure}
    \begin{subfigure}{0.7\textwidth}{\includegraphics[width=\textwidth]{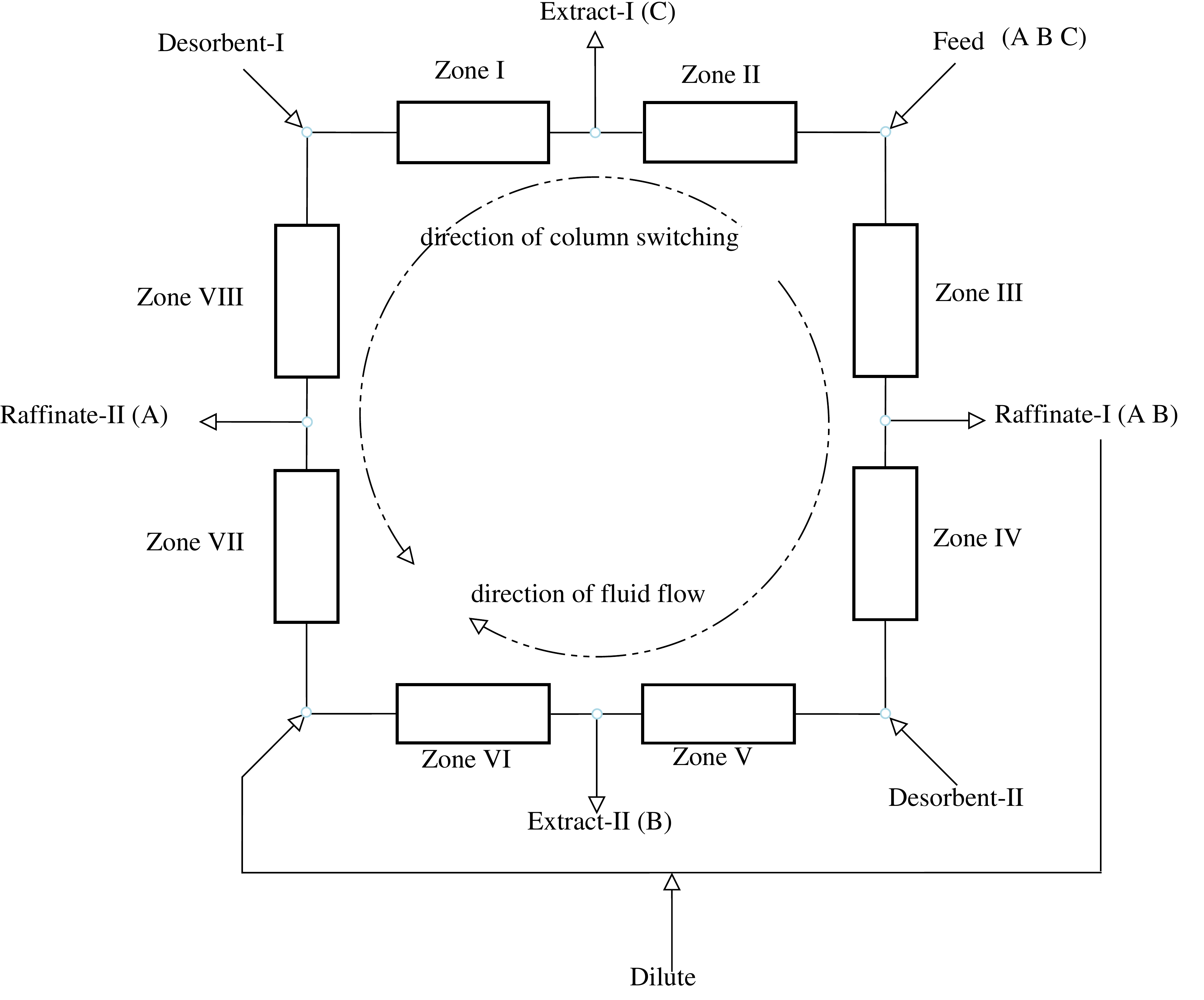}} \caption{} \label{} \end{subfigure}
    \caption{Schematics of the cascade scheme with two four-zone \textsc{iex-smb} units (top) and the integrated eight-zone scheme (bottom) applied in this study for ternary separations. In each zone, there can be multiple columns.}
    \label{fig:ternary}
\end{figure}

A cascade of two four-zone units (named as $U_1$ and $U_2$ in the following) and an integrated eight-zone scheme are applied.
The two sub-units of the cascade schemes can be connected via either the raffinate or the extract ports; while in the eight-zone schemes, either the raffinate or the extract streams of the first sub-unit can be internally bypassed.
In the cascade scheme, the flow between the raffinate stream of $U_1$ and feed stream of $U_2$ is defined as a bypass stream; while in the eight-zone scheme, the connection between the raffinate-I of the first sub-unit and the feed-II of the second sub-unit is defined as a bypass stream.
This selection depends on the equilibrium constants of the model proteins; as $k_{\text{eq}, i} = \sbk{7.70,\, 1.59,\, 35.5} \cdot 10^{-3},\, i \in \{\text{RNase, cyt, lyz}\}$, thus
\begin{equation}
    \delta\pbk{\abs{k_\text{eq, RNase} - k_\text{eq, cyt}} }< \delta\pbk{\abs{k_\text{eq, lyz} - k_\text{eq, cyt}}}
\end{equation}
indicates we should choose the configurations (as shown in Fig.~\ref{fig:ternary}) that firstly separate the strongest adsorbed component, lyz, at the extract port of the first sub-unit in both schemes, and the two sub-units are connected via the raffinate of the first sub-unit.
Thereafter, the binary separation of the target component, cyt, and the weakest adsorbed component, RNase, are achieved in the second sub-units.
In both schemes, three single columns in each zone and 24 columns at all are designed.
% and argue why this is appropriate.

%In this study, the volumetric flowrate of zone I is defined as recycle flowrate, $Q^{\text{rec}}$.
In each sub-unit of the cascade scheme, the optimized operating parameters are the switching time (same for both units as they are synchronously switched), recycle flowrate, the inlet and outlet flowrates, and the salt concentrations at the inlet ports, $\theta = \{t_s, \sbk{Q^I}_{U_1}, \sbk{Q^D}_{U_1},\allowbreak \sbk{Q^E}_{U_1}, \sbk{Q^F}_{U_1}, \sbk{c_0^F}_{U_1},\allowbreak \sbk{c_0^D}_{U_1}, \sbk{Q^I}_{U_2}, \sbk{Q^D}_{U_2},\allowbreak \sbk{Q^E}_{U_2}, \sbk{Q^F}_{U_2},\allowbreak \sbk{c_0^F}_{U_2}, \sbk{c_0^D}_{U_2}\}$.
While in the eight-zone scheme, the switch time, the recycle flowrate, the inlet and outlet flowrates, and salt concentrations at inlet ports are optimized,
that is, $\theta = \cbk{t_s, Q^I, Q^{D1},\allowbreak Q^{E1}, Q^{F1}, Q^{R1},\allowbreak Q^{D2}, Q^{E2}, Q^{F2}, c_0^{F1},\allowbreak c_0^{F2}, c_0^{D1}, c_0^{D2}}$.
The flowrates $Q^R$ and $Q^{R2}$ are calculated from flowrate balances.
%\begin{equation}
    %Q^{R2} = Q^{D1} + Q^{D2} + Q^{F1} + Q^{F2} - Q^{R1} - Q^{E1} - Q^{E2}.
%\end{equation}

\section{Results and discussion}

\subsection{Single column}
\begin{figure}[h]
    \centering
    \includegraphics[width=0.7\textwidth]{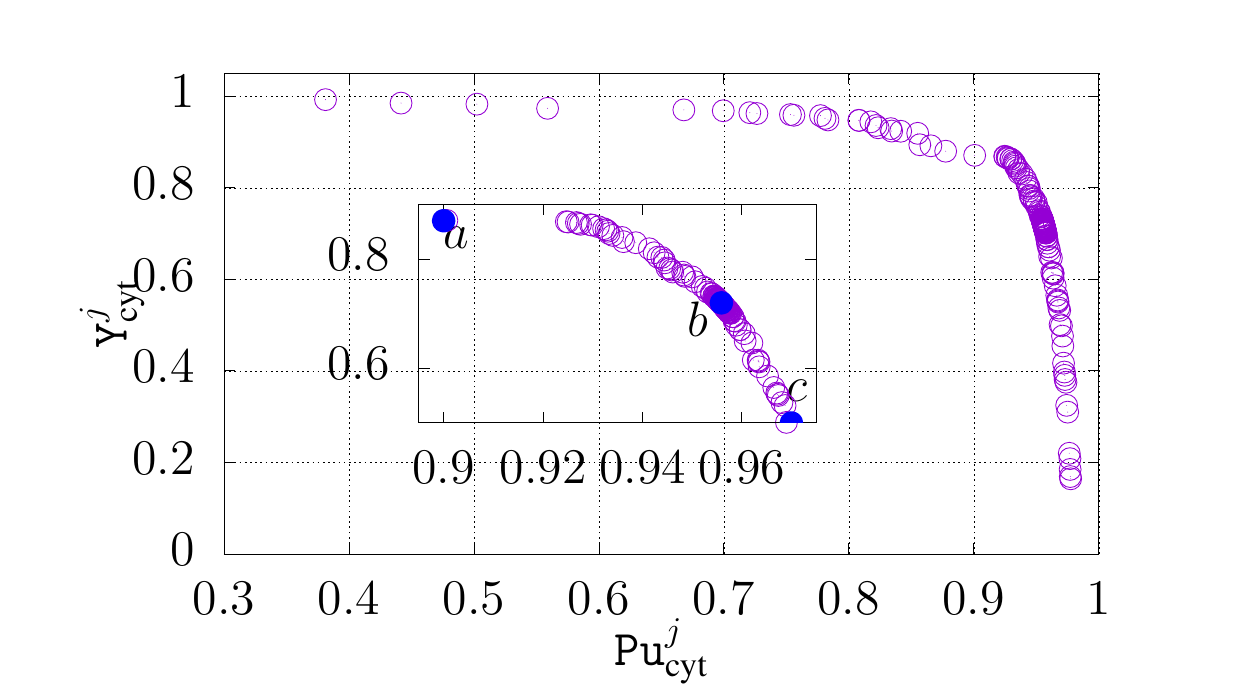} 
    \caption{Pareto optimal front of the yield and purity performance indicators in the single column system, where $a, b, c$ are three characteristic points taken from the Pareto front to show the resulting chromatograms.}
    \label{fig:pareto}
\end{figure}

The gradient shapes of the single column process are first optimized in order to have a reference for the \textsc{iex-smb} processes.
Search domain of the parameters, $\theta$, is listed in Tab.~\ref{tab:bounds_batch}.
The intervals are based on literature values with additional safety margins.
The maximal sampling size here is $\num{1.2e4}$ and the \emph{burn-in} length is set to $50\%$ of the samples.
\begin{table}
    \centering
    \scriptsize
    \caption{Search domain for the process design of the single column system.}
    \label{tab:bounds_batch}
    \begin{tabular}{c l S S S}
        \toprule
        {\multirow{2}{*}{Symbol}}   & {\multirow{2}{*}{Description}}    & \multicolumn{2}{c}{{Value}}   & {\multirow{2}{*}{Unit}} \\
        \cline{3-4}
                                    &                                   &   {min}   & {max}             & \\ 
        \midrule                                                        
        $\Delta t_1$                & elution interval one              & 500       & 8000              & \si{\second} \\
        $\Delta t_2$                & elution interval two              & 1000      & 8000              & \si{\second} \\
        $m_1$                       & elution gradient one              & 1.0e-3    & 1.0e-2            &  \\
        $m_2$                       & elution gradient two              & 1.0e-3    & 20                &  \\
        $c_{\text{init},0}$         & initial salt conc.                & 20        & 200               & \si{\mole\per\cubic\metre} \\
        \bottomrule
    \end{tabular}
\end{table}
Fig.~\ref{fig:pareto} shows the Pareto optimal front between purity and yield.
At a rather low purity requirement of $85\%$, a yield of 0.9 can be achieved.
However, at purity of $98\%$, the yield drops dramatically to 0.1.
The Pareto front provides full trade-off information of the single column process; the corresponding operating conditions, that render the Pareto optima on demand in application, can be chosen on purpose.

\begin{figure}
    \centering
    \begin{subfigure}{0.6\textwidth}{\includegraphics[width=\textwidth]{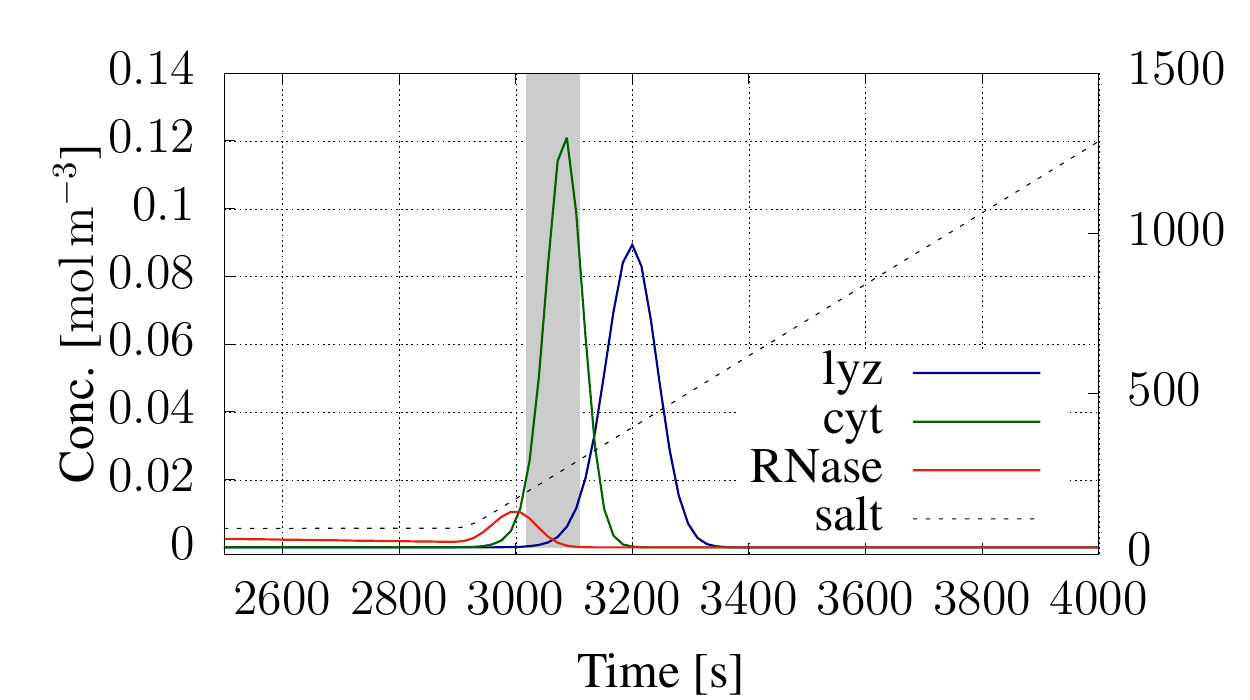}} \caption{\mbox{Point $a$; $\mathcal{PI}_{\text{cyt}}^B = [90.0\%, 0.85]$}} \label{fig:batch_demo1} \end{subfigure}
    \begin{subfigure}{0.6\textwidth}{\includegraphics[width=\textwidth]{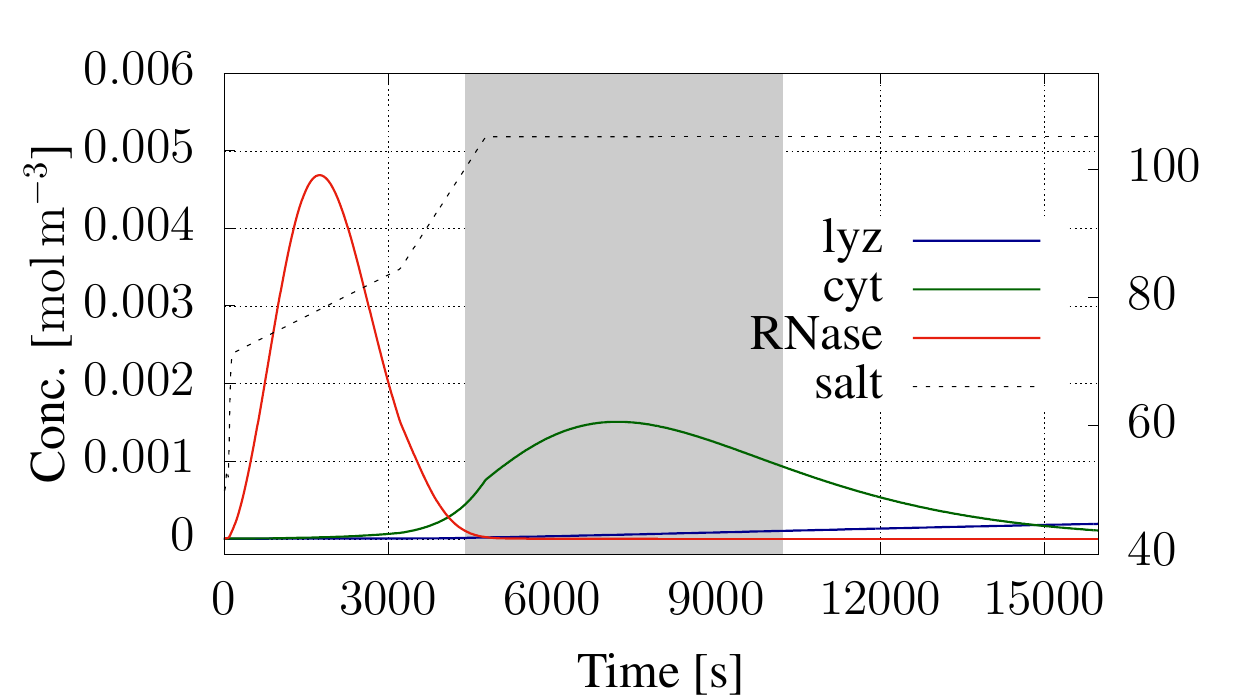}} \caption{\mbox{Point $b$; $\mathcal{PI}_{\text{cyt}}^B = [95.59\%, 0.71]$}} \label{fig:batch} \end{subfigure}
    \begin{subfigure}{0.6\textwidth}{\includegraphics[width=\textwidth]{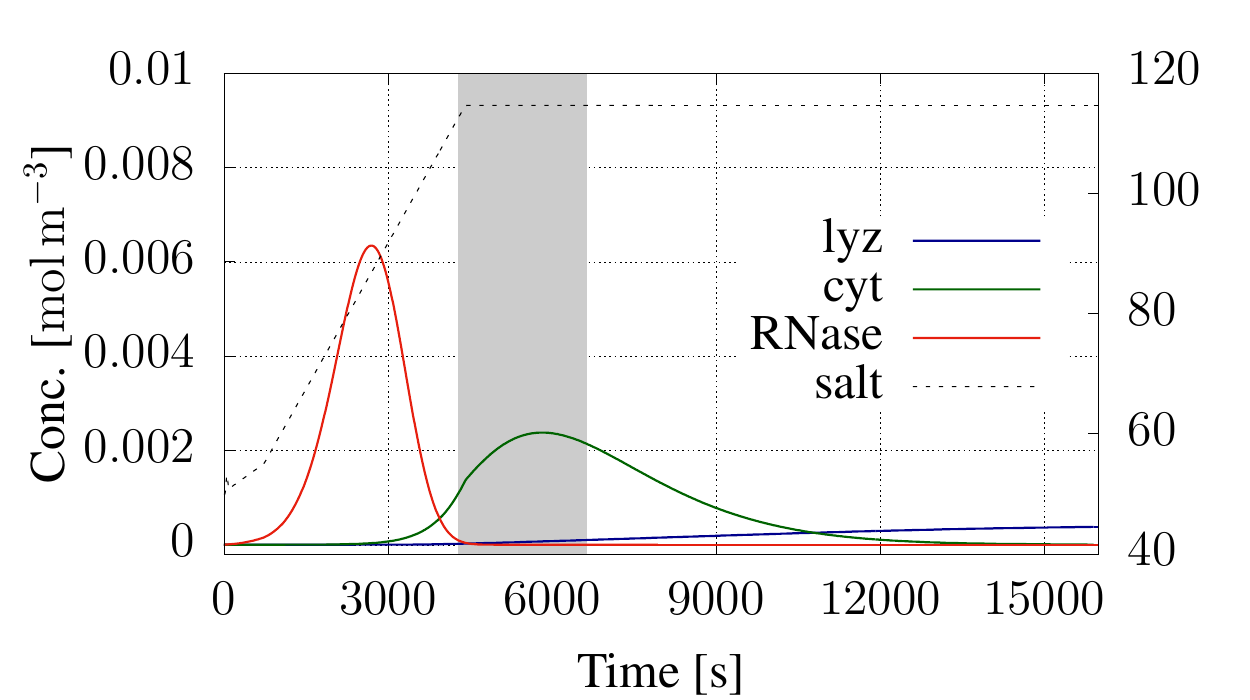}} \caption{\mbox{Point $c$; $\mathcal{PI}_{\text{cyt}}^B = [97.02\%, 0.49]$}} \label{fig:batch_demo3} \end{subfigure}
    \caption{Chromatograms of the single column system (solid lines) and the corresponding multi-step elution gradients (dashed lines).}
    \label{fig:batch_demo}
\end{figure}

Three characteristic points (\ie, $a, b, c$) on the Pareto front are exemplified and compared.
In Fig.~\ref{fig:batch_demo}, the corresponding chromatograms and multi-step salt gradients are shown; the gray areas indicate the pooling time intervals, $t_p$, of the target component.
The calculated performance indicators of the three characteristic points are as follows:
$\mathcal{PI}_{\text{cyt}}^B = \sbk{\mathtt{Pu}_{\text{cyt}}^B,\, \mathtt{Y}_{\text{cyt}}^B}$ is $[90.0\%,\, 0.85]$, and the productivity is $\mathtt{Pr}_{\text{cyt}}^B = \SI{4.94e-3}{\mole\per\cubic\metre\per\second}$ of the point $a$ (see Fig.~\ref{fig:batch_demo1}).
While they are $[95.59\%,\, 0.71]$ and \SI{7.98e-5}{\mole\per\cubic\metre\per\second} of the point $b$ (see Fig.~\ref{fig:batch}), $[97.02\%,\, 0.49]$ and \SI{1.36e-4}{\mole\per\cubic\metre\per\second} of the point $c$ (see Fig.~\ref{fig:batch_demo3}). 
The respective operating parameters are listed in Tab.~\ref{tab:est_batch}.

\begin{table}
    \centering
    \scriptsize
    \caption{Operating parameters of points $a, b, c$ in the single column system.}
    \label{tab:est_batch}
    \begin{tabular}{c S S S S}
        \toprule
        $\theta$            & $a$       & $b$       & $c$ \\ 
        \midrule                                                
        $\Delta t_1$        & 2.81e3    & 3.12e3    & 6.26e2    & \si{\second} \\
        $\Delta t_2$        & 3.53e3    & 1.56e3    & 3.68e3    & \si{\second} \\
        $m_1$               & 1.28e-3   & 4.29e-3   & 6.47e-3   &  \\
        $m_2$               & 1.11      & 1.32e-2   & 1.62e-2   &  \\
        $c_{\text{init},0}$ & 77.2      & 71.7      & 51.0      & \si{\mole\per\cubic\metre} \\
        \bottomrule
    \end{tabular}
\end{table}

\begin{figure}[h]
    \centering
    \includegraphics[width=0.6\textwidth]{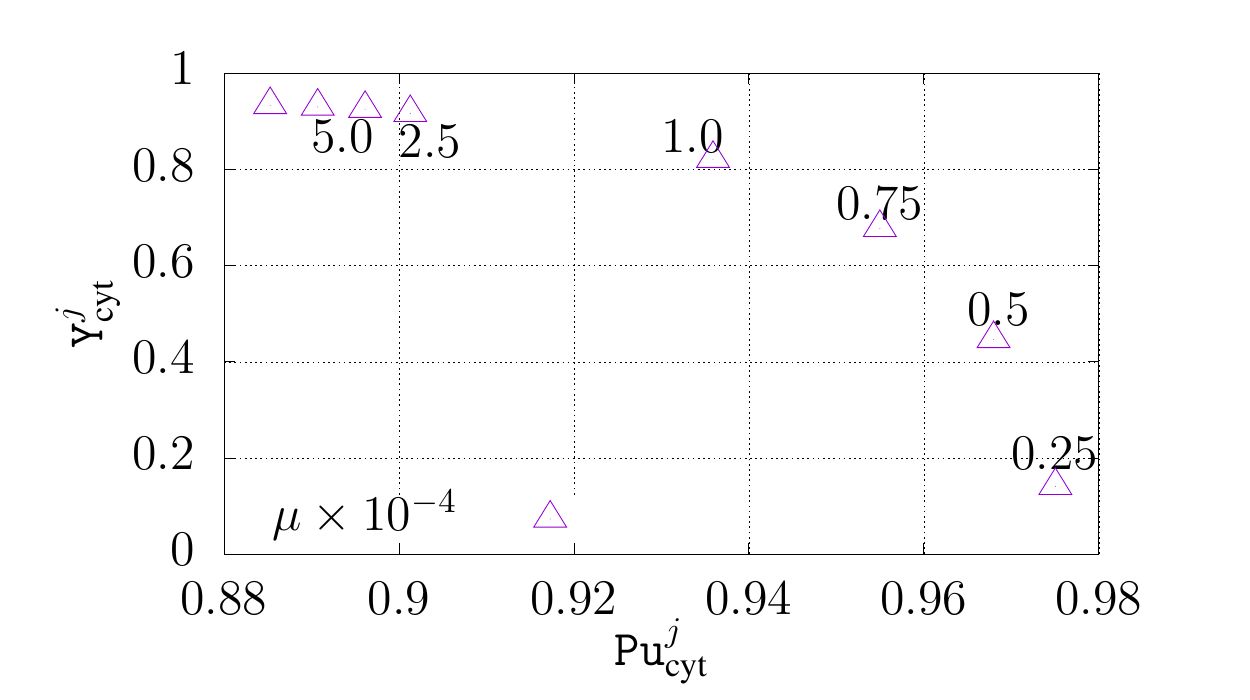}
    \caption{Impact of $\mu$ values on the trade-off relationship of purity and yield. $\mu$ is decreased from $\num{1e-4}$ to $\num{2.5e-5}$ with equivalent gap of $\num{2.5e-5}$.}
    \label{fig:pareto_eps}
\end{figure}

The pooling time length $t_p = \SI{5808}{\second}$ in Fig.~\ref{fig:batch} is much longer than that in Fig.~\ref{fig:batch_demo1} (\ie, \SI{112}{\second}) and Fig.~\ref{fig:batch_demo3} (\ie, \SI{2352}{\second}). 
In this study, pooling time intervals were calculated such that the concentrations of other two components (\ie, RNase and lyz) are lower than a predefined threshold, $\mu = \num{7.5e-5}$.
%The selection of $\mu$ definitely has an impact on the shape of the Pareto front.
Based on the same operating parameters, if we set $\mu$ to a larger value, higher yield and lower purity will be obtained.
Therefore, the selection is also a trade-off, which is illustrated in a Pareto front in Fig.~\ref{fig:pareto_eps}, where $\mu$ was decreased from \num{1e-4} to \num{2.5e-5} with equivalent gap of $\num{2.5e-5}$ using operating parameters listed in the $b$ column (\ie, point $b$) of Tab.~\ref{tab:est_batch}.

Shorter pooling time interval causes the cyt peak to be more concentrated.
This can result in high yield but larger overlap areas (thus low purity), see Fig.~\ref{fig:batch_demo1}.
When pooling time interval is larger, the cyt peak is more spread and it renders relatively lower yield but higher purity, see Fig.~\ref{fig:batch}.
As observed in Fig.~\ref{fig:batch_demo}, the peak of cyt always overlaps with the other two peaks, base line separation is not achieved.
In Fig.~\ref{fig:batch} and \ref{fig:batch_demo3}, even when lyz is hardly washed out, it begins to be eluted early at small salt concentrations.
Additionally, as lyz is almost not eluted out in Fig.~\ref{fig:batch} and \ref{fig:batch_demo3}, a strip step of lyz is needed before the regeneration step, in order to make the column reusable.
%Because of the overlaps, it is hard to design a process that entirely separates cyt from the other two components.
These overlaps may indicate that cyt can not be collected with $100\%$ purity with the single column batch system.

\subsection{Cascade scheme}
\begin{figure}
    \centering
    \includegraphics[width=0.6\textwidth]{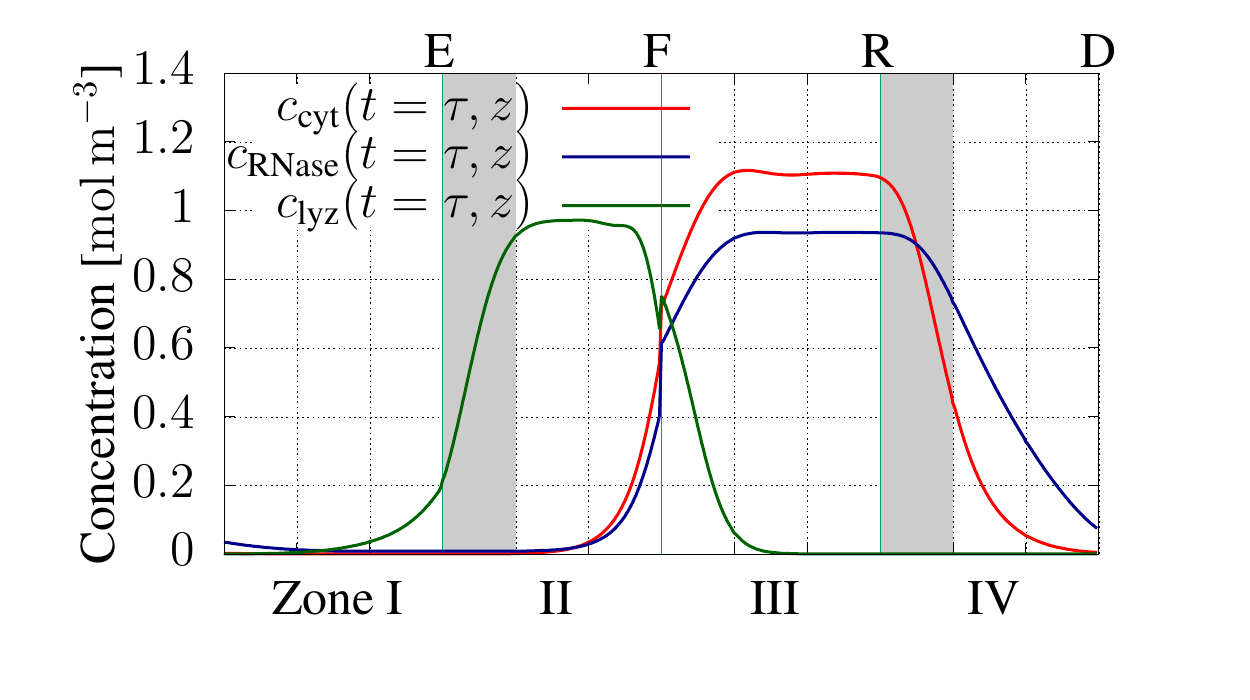}
    \caption{Chromatograms of the first sub-unit in the cascade \textsc{iex-smb} upon \textsc{css} from the empirical design.}
    \label{fig:cascade_unit1_emp}
\end{figure}

Numerical optimization of the \textsc{iex-smb} schemes require suitable initial values.
To this end, empirical rules have been developed with big manual efforts to achieve the center-cut separation first.
With the operating conditions listed in the \emph{empirical} column of Tab.~\ref{tab:cascade_op}, lyz is collected at the R port of $U_1$ with $\sbk{\mathcal{PI}_{\text{lyz}}^R}_{U_1} = [98.23\%,\, 0.99]$ (see Fig.~\ref{fig:cascade_unit1_emp}). 
In $U_2$, cyt spreads towards the R port, resulting low yield at the E port (\ie, $\sbk{\mathtt{Y}_{\text{cyt}}^E}_{U_2} = 0.33$) and low purity at the R port (\ie, $\sbk{\mathtt{Pu}_{\text{RNase}}^R}_{U_2} = 58.39\%$) (the chromatogram is not shown).
%As described in the introduction, the modifier, salt, plays an important role in the \textsc{iex-smb} chromatography and imposes challenges on the designs.
%The information gained in the empirical design is used as prior information and numerical optimization is then further applied.
Based on the initial values with safety margins, search domain as shown in Tab.~\ref{tab:bounds_cascade} is used for numerical optimization.

%\begin{figure}
    %\centering
    %\begin{subfigure}[b]{0.49\textwidth}{\includegraphics[width=\textwidth]{cascade_chroma_unit1}} \caption{} \label{fig:cascade_unit1_chroma} \end{subfigure}
    %\begin{subfigure}[b]{0.49\textwidth}{\includegraphics[width=\textwidth]{cascade_salt_unit1}} \caption{} \label{fig:cascade_unit1_salt} \end{subfigure}
    %\caption{Left: Chromatograms of the first sub-unit in the cascade \textsc{iex-smb} upon \textsc{css} from the empirical design (left). Right: Salt profiles, $c_0(t, z)$, along the columns of the first sub-unit upon the \textsc{css} ($t = \tau$)}.
    %\label{fig:cascade_unit1_emp}
%\end{figure}

\begin{figure}
    \centering
    \includegraphics[width=0.7\textwidth]{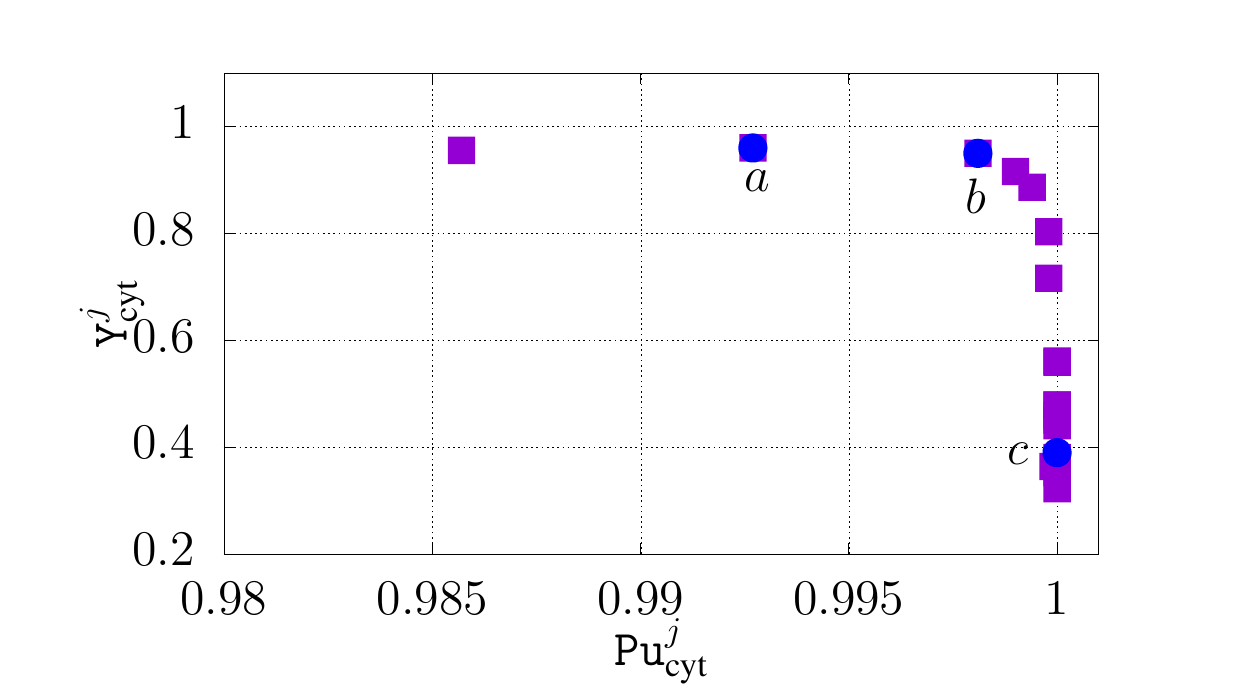}
    \caption{Pareto front of the purity and yield performance indicators of the cascade scheme from numerical optimization, where $a, b, c$ are three characteristic points taken from the Pareto front to show the resulting chromatograms.}
    \label{fig:pareto_cascade}
\end{figure}

\textsc{smb} processes often undergo a long ramp-up phase first and eventually enter into a \textsc{css}.
Elaborately choosing the initial state of columns can also accelerate the convergence \citep{bentley2014experimental}.
In this case, a total of ca.~$k = 108$ switches was required for each \textsc{iex-smb} simulation to fall below the tolerance error of $e_t = \num{1e-5}$, see Eq.~\eqref{eq:difference_method}.
Thus, gaining points for generating Pareto fronts in \textsc{smb} chromatographic processes is computational expensive.
\cite{li2014using} have proposed computationally cheap surrogate models for efficient optimization of \textsc{smb} chromatography.
In this study, only the operating parameters of $U_2$ are optimized; the operating conditions of $U_1$ are taken directly from the empirical design.
Moreover, only the Pareto optimal fragment with purity higher than $99\%$ is concerned, such that fewer points are requested.
The maximal sampling size of \textsc{mcmc} is $\num{300}$, and the \emph{burn-in} length is $\num{50}$.
%The Pareto front of the cascade scheme is illustrated in .
As seen from the Pareto fronts in Fig.~\ref{fig:pareto_cascade}, the cascade scheme outperforms the single column system in both purity and yield.

\begin{figure}
    \centering
    %\begin{subfigure}[b]{0.49\textwidth}{\includegraphics[width=\textwidth]{cascade_chroma_unit1}} \caption{$U_1$} \label{fig:cascade_chroma_0} \end{subfigure}
    \begin{subfigure}[b]{0.7\textwidth}{\includegraphics[width=\textwidth]{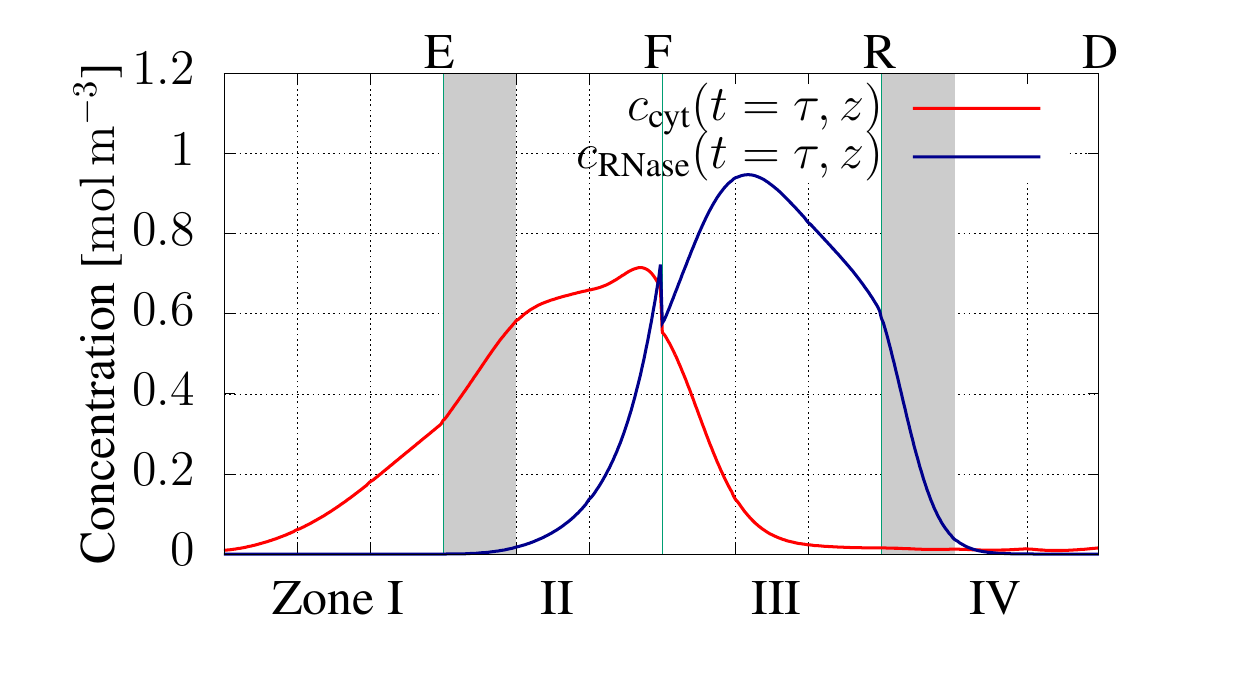}} \caption{\mbox{Point $a$; $\mathcal{PI}_{\text{cyt}}^E = [99.27\%, 0.96], \mathtt{Pr}_{\text{cyt}}^E = \SI{3.13e-4}{\mole\per\cubic\metre\per\second}$}} \label{fig:cascade_chroma_1}  \end{subfigure}
    \begin{subfigure}[b]{0.7\textwidth}{\includegraphics[width=\textwidth]{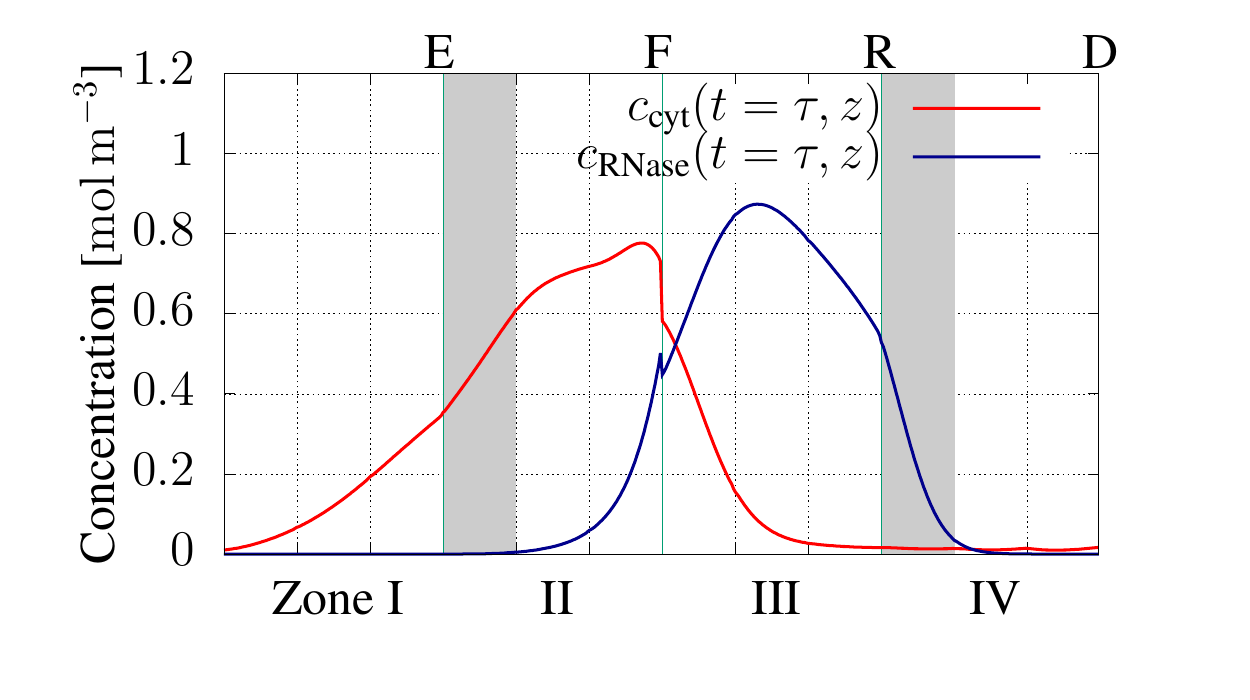}} \caption{\mbox{Point $b$; $\mathcal{PI}_{\text{cyt}}^E = [99.81\%, 0.95], \mathtt{Pr}_{\text{cyt}}^E = \SI{3.15e-4}{\mole\per\cubic\metre\per\second}$}} \label{fig:cascade_chroma_2} \end{subfigure}
    \begin{subfigure}[b]{0.7\textwidth}{\includegraphics[width=\textwidth]{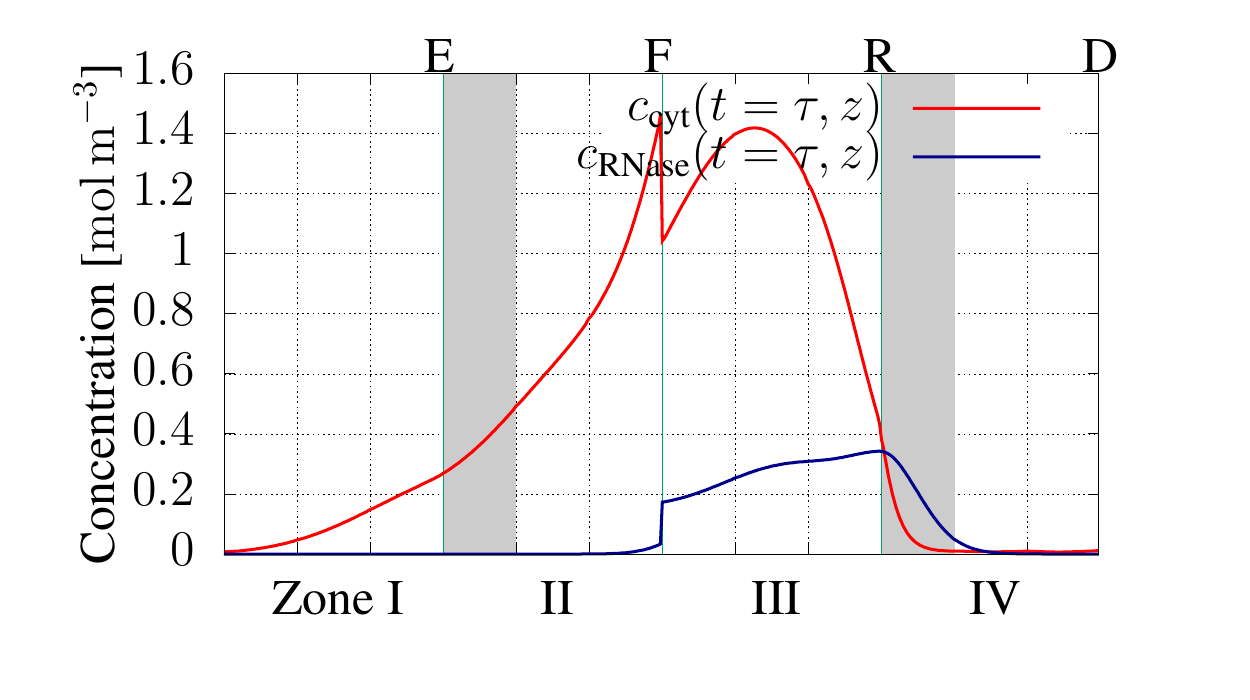}} \caption{\mbox{Point $c$; $\mathcal{PI}_{\text{cyt}}^E = [100\%, 0.39], \mathtt{Pr}_{\text{cyt}}^E = \SI{1.30e-4}{\mole\per\cubic\metre\per\second}$}} \label{fig:cascade_chroma_3} \end{subfigure}
    \caption{Chromatograms of the second sub-unit in the cascade \textsc{iex-smb} upon \textsc{css}.}% Lyz is collected at the E node of $U_1$, while cyt at E node of $U_2$ and RNase at R node of $U_2$.} %, \num{3.31e-4}, \num{1.07e-4}, \num{3.19e-4}
    \label{fig:cascade_chroma_op}
\end{figure}

Three characteristic points (\ie, $a, b, c$), ranging yields from high to low, on the Pareto front are compared. 
The corresponding operating conditions are listed in Tab.~\ref{tab:cascade_op}; and the resulting chromatograms are shown in Fig.~\ref{fig:cascade_chroma_op}.
Combined axial concentration profiles in the columns are displayed at multiples of the switching time.
The performance vector $\sbk{\mathcal{PI}_{\text{cyt}}^E}_{U_2} = \sbk{\mathtt{Pu}^E_{\text{cyt}},\, \mathtt{Y}_{\text{cyt}}^E}_{U_2}$ of the point $a$ is $[99.27\%,\, 0.96]$ (cf.~Fig.~\ref{fig:cascade_chroma_1}), $[99.81\%,\, 0.95]$ of the point $b$ (cf.~Fig.~\ref{fig:cascade_chroma_2}) and $[100\%,\, 0.39]$ of the point $c$ (cf.~Fig.~\ref{fig:cascade_chroma_3}).
After numerical optimization, the performance indicators tremendously increase from that of the empirical design, $[97.92\%,\, 0.33]$.
The main difference between the chromatograms of $a$ and $b$ is that less RNase keeps retained to the port E in $b$, resulting in higher purity at the port E.
The productivities of cyt, $\sbk{\mathtt{Pr}_{\text{cyt}}^E}_{U_2}$, are $[\num{3.13e-4}, \num{3.15e-4},\allowbreak \num{1.30e-4}]\, \si{\mole\per\cubic\metre\per\second}$, respectively.
It indicates that productivity is high when the concentration values in the shadow areas at the port E are high and flat.
Though RNase and lyz are treated as impurities in the center-cut separation, they can be withdrawn with high purity and yield in this study.
For instance, $\sbk{\mathcal{PI}_{\text{RNase}}^R}_{U_2} = [96.14\%,\, 0.97]$ of the point $a$, $[95.45\%,\, 0.98]$ of the point $b$.
However, as pointed out by \cite{nicolaos2001application}, it is not impossible to achieve three pure outlets with both the cascade and the integrated schemes.
%while the indicator of RNase at the raffinate port of $U_2$ from $\mathcal{PI}_{\text{RNase}}^R= [\mathtt{Pu}^R_{\text{RNase}}, \mathtt{Y}_{\text{RNase}}^R] = [58.39\%, 0.71]$ of the empirical design to $[95.45\%, 0.98]$.

\begin{figure}
    \centering
    \begin{subfigure}[b]{0.49\textwidth}{\includegraphics[width=\textwidth]{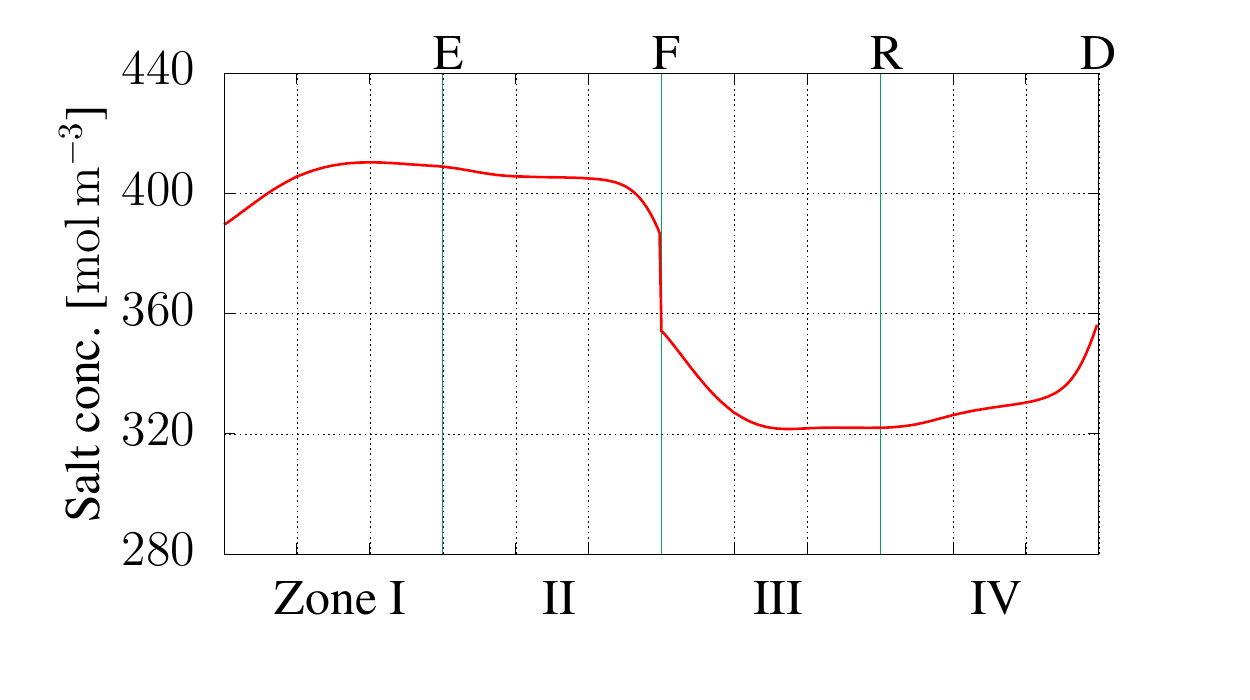}} \caption{$U_1$} \label{fig:cascade_salt_u1} \end{subfigure}
    \begin{subfigure}[b]{0.49\textwidth}{\includegraphics[width=\textwidth]{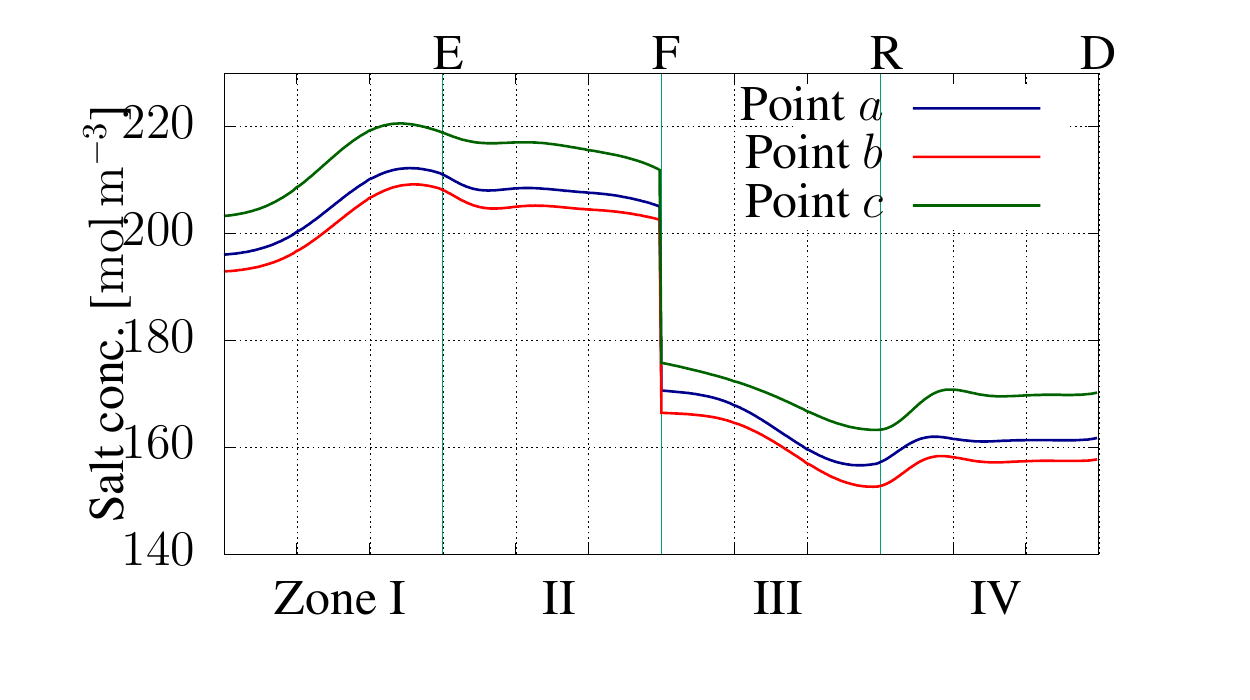}} \caption{$U_2$} \label{fig:cascade_salt_u2} \end{subfigure}
    \caption{Salt profiles, $c_0(t, z)$, along the columns of the cascade scheme upon the \textsc{css} ($t = \tau$).}
    \label{fig:cascade_salt}
\end{figure}

The salt profiles, $c_0(t, z)$, along the two \textsc{iex-smb} units at time upon \textsc{css}, $t = \tau$, are depicted in Fig.~\ref{fig:cascade_salt}.
In $U_1$, the salt concentrations at inlet ports, $\sbk{c_0^D,\, c_0^F}_{U_1}$, used for constructing the two solvent gradients are $[290,\, 420]\, \si{\mole\per\cubic\metre}$ (cf.~Fig.~\ref{fig:cascade_salt_u1}).
The salt concentrations at the inlet ports of $U_2$, $\sbk{c_0^D,\, c_0^F}_{U_2}$, are $[206.56,\, 243.53]\, \si{\mole\per\cubic\metre}$ of the point $a$; $[203.92,\, 244.93]\,\allowbreak \si{\mole\per\cubic\metre}$ of the point $b$ and $[208,\, 243]\, \si{\mole\per\cubic\metre}$ of the point $c$ (cf.~Fig.~\ref{fig:cascade_salt_u2}).

The salt gap of solvent between zones I,II and zones III,IV of $U_1$ facilitates to firstly separate the most strongly adsorbed lyz from the ternary mixture; RNase and cyt are still not rather separated in $U_2$.
Because of the salt profile in zones III and IV (\ie, ca.~\SI{320}{\mole\per\cubic\metre}), RNase and cyt have approximately the same electrostatic interaction forces with the stationary phase, causing rather similar desorption rates off the stationary phase.
In order to separate RNase and cyt in $U_2$, the salt concentrations at the inlet ports of $U_2$ must be reduced, leading one component (\ie, cyt) to be more strongly bounded.
This can be achieved by diluting the bypass stream with pure buffer:
\begin{equation}
    \sbk{ c_{\text{in},i}^F }_{U_2} = \frac{\sbk{ Q^R c_{\text{out},i}^R }_{U_1}}{\sbk{Q^R}_{U_1} + Q^{\text{dilute}}} \qquad i \in \{0, 1, \dots, M\}
    \label{eq:dilute}
\end{equation}
In the numerical optimization, $\sbk{Q^F}_{U_2}$ and $\sbk{Q^R}_{U_1}$ are optimized, $Q^{\text{dilute}} = \sbk{Q^F}_{U_2} - \sbk{Q^R}_{U_1}$ and $\sbk{c_{\text{in},i}^F}_{U_2}$ are subsequently changed.
As shown in Fig.~\ref{fig:cascade_salt_u2}, the salt concentration in $U_2$ is decreased to ca.~\SI{200}{\mole\per\cubic\metre} in zones I,II and \SI{160}{\mole\per\cubic\meter} in zones III,IV.
With these salt concentrations, RNase and cyt can be separated in $U_2$.
With the dilution, the protein concentrations are decreased as a side effect.
Therefore, the peaks in $U_2$ are not as concentrated as observed in $U_1$;
they distribute over several zones, which makes the process design of the $U_2$ challenging.

\begin{landscape}

    \begin{table}
        \centering
        \scriptsize
        \caption{Operating parameters of the point $a, b, c$ in the cascade \textsc{iex-smb} units.}
        \label{tab:cascade_op}
        \begin{tabular}{c l S S S S S c}
            \toprule
            \multirow{2}{*}{Symbol} & \multirow{2}{*}{Description} & \multicolumn{2}{c}{$\text{Empirical}$} & \multicolumn{3}{c}{$\text{Numerical}$}     & \multirow{2}{*}{Unit} \\
            \cline{3-4} \cline{5-7}
            &                                                      & {$U_1$}  & {$U_2$}   & $a$     & $b$     & $c$ & \\
            \midrule                                                                                                    
            $Q^F$                   & feed flowrate                & 0.55e-8  & 0.75e-8   & 1.48e-8 & 1.37e-8 & 1.56e-8    & \si{\cubic\metre\per\second} \\
            $Q^R$                   & raffinate flowrate           & 0.60e-8  & 0.90e-8   & 1.49e-8 & 1.36e-8 & 1.63e-8    & \si{\cubic\metre\per\second} \\
            $Q^D$                   & desorbent flowrate           & 1.14e-8  & 1.34e-8   & 1.24e-8 & 1.22e-8 & 1.24e-8    & \si{\cubic\metre\per\second} \\
            $Q^E$                   & extract flowrate             & 1.09e-8  & 1.19e-8   & 1.23e-8 & 1.22e-8 & 1.17e-8    & \si{\cubic\metre\per\second} \\
            $Q^{\text{I}}$          & zone I flowrate              & 2.21e-8  & 2.56e-8   & 2.96e-8 & 2.90e-8 & 3.01e-8    & \si{\cubic\metre\per\second} \\
            $Q^{\text{II}}$         & zone II flowrate             & 1.12e-8  & 1.37e-8   & 1.73e-8 & 1.63e-8 & 1.84e-8    & \si{\cubic\metre\per\second} \\
            $Q^{\text{III}}$        & zone III flowrate            & 1.67e-8  & 2.12e-8   & 3.21e-8 & 3.04e-8 & 3.40e-8    & \si{\cubic\metre\per\second} \\
            $Q^{\text{IV}}$         & zone IV flowrate             & 1.07e-8  & 1.22e-8   & 1.72e-8 & 1.69e-8 & 1.77e-8    & \si{\cubic\metre\per\second} \\
            $c^F_0$                 & feed salt conc.              & 290      & 200       & 206.56  & 203.92  & 208        & \si{\mole\per\cubic\metre} \\
            $c^D_0$                 & desorbent salt conc.         & 420      & 240       & 243.53  & 244.93  & 243        & \si{\mole\per\cubic\metre} \\
            \midrule
            $t_s$                   & switching time               & \multicolumn{5}{c}{100}       & \si{\second} \\
            \bottomrule
        \end{tabular}
    \end{table}

    \begin{table}
        \centering
        \scriptsize
        \caption{Boundary conditions of the operating parameters of the cascade scheme.}
        \label{tab:bounds_cascade}
        \begin{tabular}{c l S S S}
            \toprule
            {\multirow{2}{*}{Symbol}}   & {\multirow{2}{*}{Description}}    & \multicolumn{2}{c}{{Value}}   & {\multirow{2}{*}{Unit}} \\
            \cline{3-4}
                                        &                                   &   {min}       & {max}         & \\ 
            \midrule                                                        
            $Q^I$                       & zone I flowrate                  & 2.06e-8       & 3.06e-8        & \si{\cubic\metre\per\second} \\
            $Q^{F}$                    & feed one flowrate                  & 0.84e-8       & 1.84e-8       & \si{\cubic\metre\per\second} \\
            $Q^{D}$                    & desorbent one flowrate             & 0.69e-8       & 1.69e-8       & \si{\cubic\metre\per\second} \\
            $Q^{E}$                    & extract one flowrate               & 0.35e-8       & 1.25e-8       & \si{\cubic\metre\per\second} \\
            $c^{F}_0$                  & feed two salt conc.                & 180           & 220           & \si{\mole\per\cubic\metre} \\
            $c^{D}_0$                  & desorbent two salt conc.           & 220           & 260           & \si{\mole\per\cubic\metre} \\
            \bottomrule
        \end{tabular}
    \end{table}

    \begin{table}
        \centering
        \scriptsize
        \caption{Operating parameters of the point $a, b, c$ in the eight-zone \textsc{iex-smb}}
        \label{tab:8-zone}
        \begin{tabular}{c l S S S S S}
            \toprule
            {\multirow{2}{*}{Symbol}}& {\multirow{2}{*}{Description}}  & \multicolumn{4}{c}{{Value}}  & {\multirow{2}{*}{Unit}} \\
            \cline{3-6}
            &                                                  & {Empirical} & $a$        & $b$        & $c$ & \\ 
            \midrule                                                                                               
            $Q^{F1}$                & feed one flowrate        & 0.477e-8    & 0.544e-8   & 0.544e-8   & 0.554e-8   & \si{\cubic\metre\per\second} \\
            $Q^{R1}$                & raffinate one flowrate   & 0.518e-8    & 0.586e-8   & 0.574e-8   & 0.558e-8   & \si{\cubic\metre\per\second} \\
            $Q^{D1}$                & desorbent one flowrate   & 1.086e-8    & 1.072e-8   & 1.082e-8   & 1.089e-8   & \si{\cubic\metre\per\second} \\
            $Q^{E1}$                & extract one flowrate     & 1.056e-8    & 1.023e-8   & 1.020e-8   & 1.045e-8   & \si{\cubic\metre\per\second} \\
            $Q^{\text{I}}$          & zone I flowrate          & 2.166e-8    & 2.000e-8   & 2.021e-8   & 2.000e-8   & \si{\cubic\metre\per\second} \\
            $Q^{\text{II}}$         & zone II flowrate         & 1.110e-8    & 0.977e-8   & 1.001e-8   & 0.955e-8   & \si{\cubic\metre\per\second} \\
            $Q^{\text{III}}$        & zone III flowrate        & 1.587e-8    & 1.521e-8   & 1.540e-8   & 1.508e-8   & \si{\cubic\metre\per\second} \\
            $Q^{\text{IV}}$         & zone IV flowrate         & 1.069e-8    & 0.935e-8   & 0.966e-8   & 0.950e-8   & \si{\cubic\metre\per\second} \\
            $Q^{F2}$                & feed two flowrate        & 0.753e-8    & 0.826e-8   & 0.837e-8   & 0.843e-8   & \si{\cubic\metre\per\second} \\
            $Q^{R2}$                & raffinate two flowrate   & 1.009e-8    & 1.122e-8   & 1.164e-8   & 1.189e-8   & \si{\cubic\metre\per\second} \\
            $Q^{D2}$                & desorbent two flowrate   & 1.510e-8    & 1.391e-8   & 1.408e-8   & 1.408e-8   & \si{\cubic\metre\per\second} \\
            $Q^{E2}$                & extract two flowrate     & 1.243e-8    & 1.102e-8   & 1.109e-8   & 1.101e-8   & \si{\cubic\metre\per\second} \\
            $Q^{\text{V}}$          & zone V flowrate          & 2.579e-8    & 2.327e-8   & 2.374e-8   & 2.358e-8   & \si{\cubic\metre\per\second} \\
            $Q^{\text{VI}}$         & zone VI flowrate         & 1.336e-8    & 1.225e-8   & 1.265e-8   & 1.257e-8   & \si{\cubic\metre\per\second} \\
            $Q^{\text{VII}}$        & zone VII flowrate        & 2.089e-8    & 2.051e-8   & 2.102e-8   & 2.100e-8   & \si{\cubic\metre\per\second} \\
            $Q^{\text{VII}}$        & zone VIII flowrate       & 1.080e-8    & 0.928e-8   & 0.939e-8   & 0.911e-8   & \si{\cubic\metre\per\second} \\
            $c^{F1}_0$              & feed one salt conc.      & 295.1       & 270.0      & 270.2      & 270.5     & \si{\mole\per\cubic\metre} \\
            $c^{D1}_0$              & desorbent one salt conc. & 422.0       & 442.0      & 440.3      & 442.7     & \si{\mole\per\cubic\metre} \\
            $c^{F2}_0$              & feed two salt conc.      & 183.1       & 211.0      & 209.4      & 216.6     & \si{\mole\per\cubic\metre} \\
            $c^{D2}_0$              & desorbent two salt conc. & 240.7       & 240.0      & 240.0      & 240.5     & \si{\mole\per\cubic\metre} \\
            \midrule                                                                                               
            $t_s$                   & switching time           & 105.9       & 109.1      & 109.8      & 109.7      & \si{\second} \\
            \bottomrule
        \end{tabular}
    \end{table}
\end{landscape}

\subsection{Eight-zone scheme}
As in the design of cascade scheme, the initial values were obtained from empirical designs with manual efforts.
With the operating conditions listed in the \emph{empirical} column of Tab.~\ref{tab:8-zone}, cyt spreads from zone VI to zone VII, RNase from zone V to VI, such that $\mathtt{Pu}_{\text{RNase}}^{R2} = 73.23\%$ and $\mathtt{Pu}_{\text{cyt}}^{E2} = 87.50\%$ (the chromatogram is not shown).
The search domain for the numerical optimization is based on the above initial values with safety margins, see Tab.~\ref{tab:bounds}.
%The dilution as described in the cascade scheme was applied to the bypass stream. 

\begin{table}
    \centering
    \scriptsize
    \caption{Search domain for the design of eight-zone scheme.}
    \label{tab:bounds}
    \begin{tabular}{c l S S S}
        \toprule
        {\multirow{2}{*}{Symbol}}   & {\multirow{2}{*}{Description}}    & \multicolumn{2}{c}{{Value}}   & {\multirow{2}{*}{Unit}} \\
        \cline{3-4}
                                    &                                   &   {min}       & {max}         & \\ 
        \midrule                                                        
        $t_s$                       & switch time                       & 90            & 110           & \si{\second} \\
        $Q^I$                       & zone I flowrate                   & 2.0e-8        & 2.5e-8        & \si{\cubic\metre\per\second} \\
        $Q^{D1}$                    & desorbent one flowrate            & 0.94e-8       & 1.14e-8       & \si{\cubic\metre\per\second} \\
        $Q^{E1}$                    & extract one flowrate              & 0.99e-8       & 1.19e-8       & \si{\cubic\metre\per\second} \\
        $Q^{F1}$                    & feed one flowrate                 & 0.45e-8       & 0.65e-8       & \si{\cubic\metre\per\second} \\
        $Q^{R1}$                    & raffinate one flowrate            & 0.50e-8       & 0.70e-8       & \si{\cubic\metre\per\second} \\
        $Q^{D2}$                    & desorbent two flowrate            & 1.30e-8       & 1.55e-8       & \si{\cubic\metre\per\second} \\
        $Q^{E2}$                    & extract two flowrate              & 1.09e-8       & 1.29e-8       & \si{\cubic\metre\per\second} \\
        $Q^{F2}$                    & feed two flowrate                 & 0.65e-8       & 0.85e-8       & \si{\cubic\metre\per\second} \\
        $c^{F1}_0$                  & feed one salt conc.               & 270           & 310           & \si{\mole\per\cubic\metre} \\
        $c^{D1}_0$                  & desorbent one salt conc.          & 410           & 450           & \si{\mole\per\cubic\metre} \\
        $c^{F2}_0$                  & feed two salt conc.               & 170           & 220           & \si{\mole\per\cubic\metre} \\
        $c^{D2}_0$                  & desorbent two salt conc.          & 240           & 260           & \si{\mole\per\cubic\metre} \\
        \bottomrule
    \end{tabular}
\end{table}

\begin{figure}
    \centering
    \includegraphics[width=0.7\textwidth]{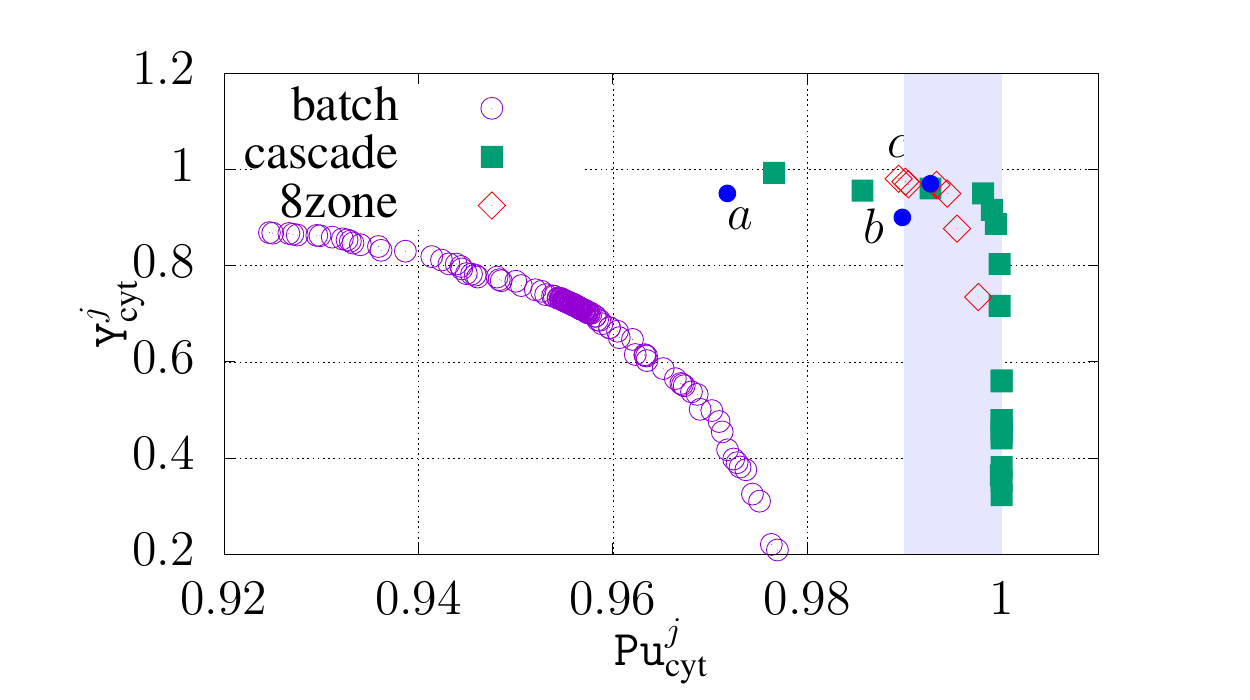}
    \caption{Pareto fronts of the purity and yield performance indicators of the eight-zone scheme from numerical optimization, where the Pareto fronts of single column and cascade scheme are superimposed. $a, b, c$ are three characteristic points taken from the Pareto front of the eight-zone scheme.}
    \label{fig:pareto_8zone}
\end{figure}

In this case, a total of ca.~$k = 148$ switches was required for each eight-zone numerical simulation to fall below the tolerance error of $e_t = \num{1e-5}$.
The convergence time from an initial state to the \textsc{css}, thus the total computational time in the eight-zone scheme, is longer than that in the cascade scheme.
%In this study, only the Pareto optimal fragment with purity higher than $99\%$ is concerned, such that fewer points can be requested.
The maximal sampling size of \textsc{mcmc} is $\num{300}$, and the \emph{burn-in} length is $\num{50}$.
The Pareto optimal front with purity larger than $99\%$ is illustrated in Fig.~\ref{fig:pareto_8zone}; the Pareto fronts of single column and cascade systems are superimposed for comparison.
The blue shadow areas show the constraint region.

\begin{figure}
    \centering
    \begin{subfigure}[b]{0.7\textwidth}{\includegraphics[width=\textwidth]{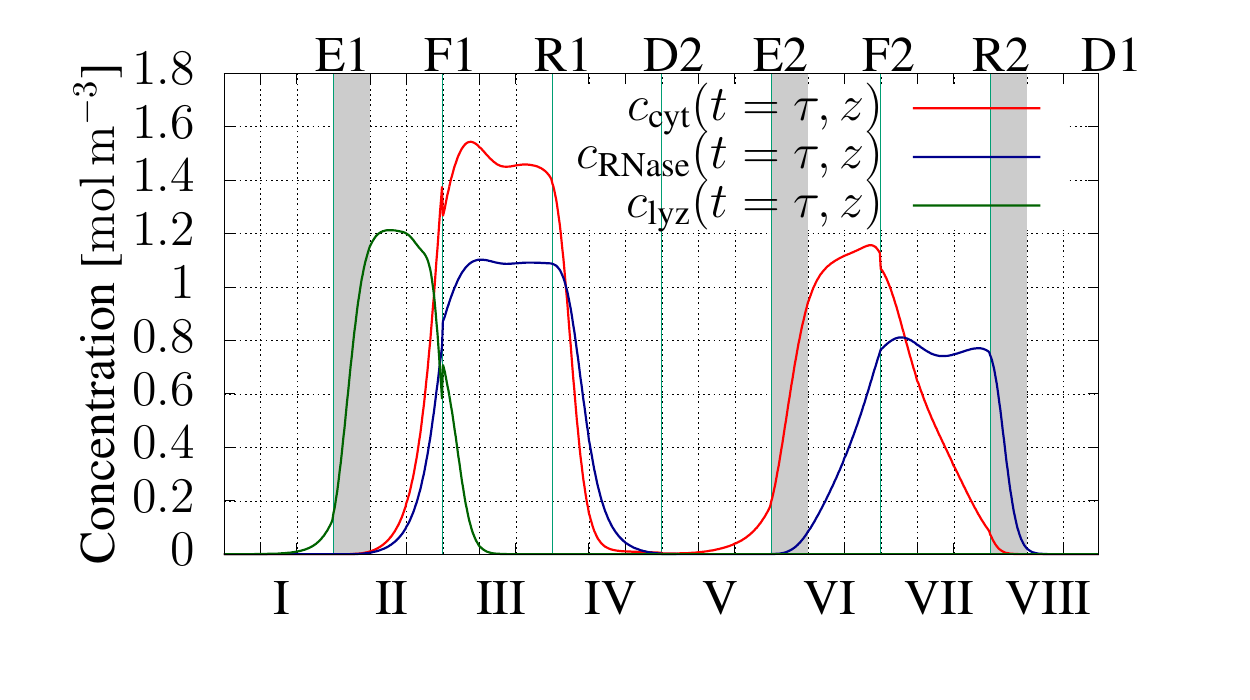}} \caption{\mbox{Point $a$; $\mathcal{PI}_{\text{cyt}}^{E2} = [97.18\%, 0.95], \mathtt{Pr}_{\text{cyt}}^{E2} = \SI{3.10e-4}{\mole\per\cubic\metre\per\second}$}} \label{fig:8-zone_chroma_opt_1} \end{subfigure}
    \begin{subfigure}[b]{0.7\textwidth}{\includegraphics[width=\textwidth]{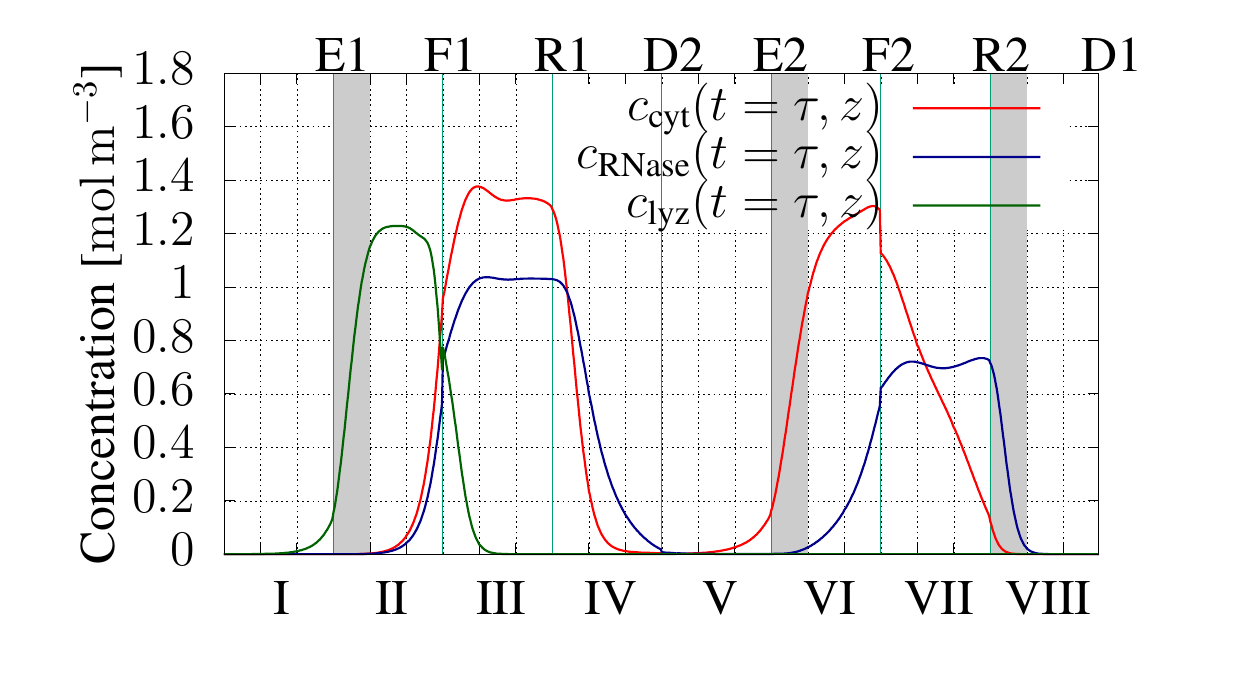}} \caption{\mbox{Point $b$; $\mathcal{PI}_{\text{cyt}}^{E2} = [98.98\%, 0.90], \mathtt{Pr}_{\text{cyt}}^{E2} = \SI{2.93e-4}{\mole\per\cubic\metre\per\second}$}} \label{fig:8-zone_chroma_opt_2} \end{subfigure}
    \begin{subfigure}[b]{0.7\textwidth}{\includegraphics[width=\textwidth]{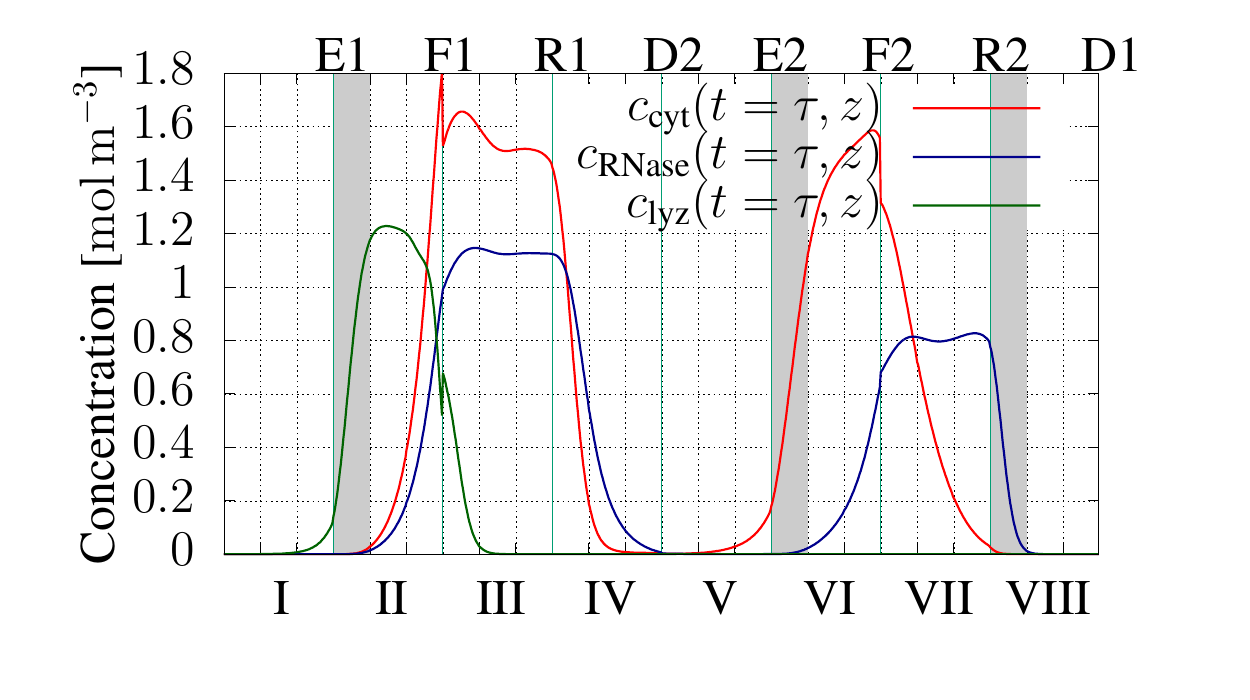}} \caption{\mbox{Point $c$; $\mathcal{PI}_{\text{cyt}}^{E2} = [99.27\%, 0.97], \mathtt{Pr}_{\text{cyt}}^{E2} = \SI{3.25e-4}{\mole\per\cubic\metre\per\second}$}} \label{fig:8-zone_chroma_opt_3} \end{subfigure}
    \caption{Chromatograms of the eight-zone \textsc{iex-smb} upon \textsc{css}.}
    \label{fig:8-zone_chroma_opt}
\end{figure}

By using the multi-objective sampling algorithm, \textsc{mcmc}, not only the Pareto front solutions but also the points near Pareto front can be studied.
In this study, three points (\ie, $a, b, c$; the point $c$ is on the Pareto front), ranging purities from low to high, were chosen to illustrate the impact of the composition of R1 stream on the performance of the second sub-unit.
The corresponding operating conditions are listed in Tab.~\ref{tab:8-zone}; and the chromatograms upon \textsc{css} are shown in Fig.~\ref{fig:8-zone_chroma_opt}. 
Combined axial concentration profiles in the columns are displayed at multiples of the switching time.
In Fig.~\ref{fig:8-zone_chroma_opt_1}, cyt is collected at the E2 node with $\mathcal{PI}_{\text{cyt}}^{E2} = [97.18\%,\, 0.95]$, while $[98.98\%,\, 0.90]$ in Fig.~\ref{fig:8-zone_chroma_opt_2} and $[99.27\%,\, 0.97]$ in Fig.~\ref{fig:8-zone_chroma_opt_3}.
As seen from Fig.~\ref{fig:8-zone_chroma_opt}, the higher concentration of the cyt composition is in the R1 stream, the higher of the yield and productivity indicators are at the E2 port.
To be specific, $\mathtt{Y}_{\text{cyt}}^{E2} = 0.95$ and $\mathtt{Pr}_{\text{cyt}}^{E2} = \SI{3.10}{\mole\per\cubic\metre\per\second}$ of the point $a$; $\mathtt{Y}_{\text{cyt}}^{E2} = 0.90$ and $\mathtt{Pr}_{\text{cyt}}^{E2} = \SI{2.93}{\mole\per\cubic\metre\per\second}$ of the point $b$, and $\mathtt{Y}_{\text{cyt}}^{E2} = 0.97$ and $\mathtt{Pr}_{\text{cyt}}^{E2} = \SI{3.25}{\mole\per\cubic\metre\per\second}$ of the point $c$.
Although in the center-cut separation, the other two components (\ie, RNase and lyz) can be viewed as impurities, they can be withdrawn with high purity and yield.
Specifically, for RNase at the R2 node it is $\mathcal{PI}_{\text{RNase}}^{R2} = [94.86\%,\, 0.97]$ of the point $a$, $[91.80\%,\, 0.99]$ of the point $b$ and $[98.29\%,\, 0.99]$ of the point $c$; 
for lyz at the E1 node it is $\mathcal{PI}_{\text{lyz}}^{E1} = [99.75\%,\, 1.00]$ of the point $a$, $[99.94\%,\, 1.00]$ of the point $b$ and $[99.31\%,\, 1.00]$ of the point $c$.
However, it is still impossible to achieve three pure outlets \citep{nicolaos2001application}.

\begin{figure}
    \centering
    \includegraphics[width=0.7\textwidth]{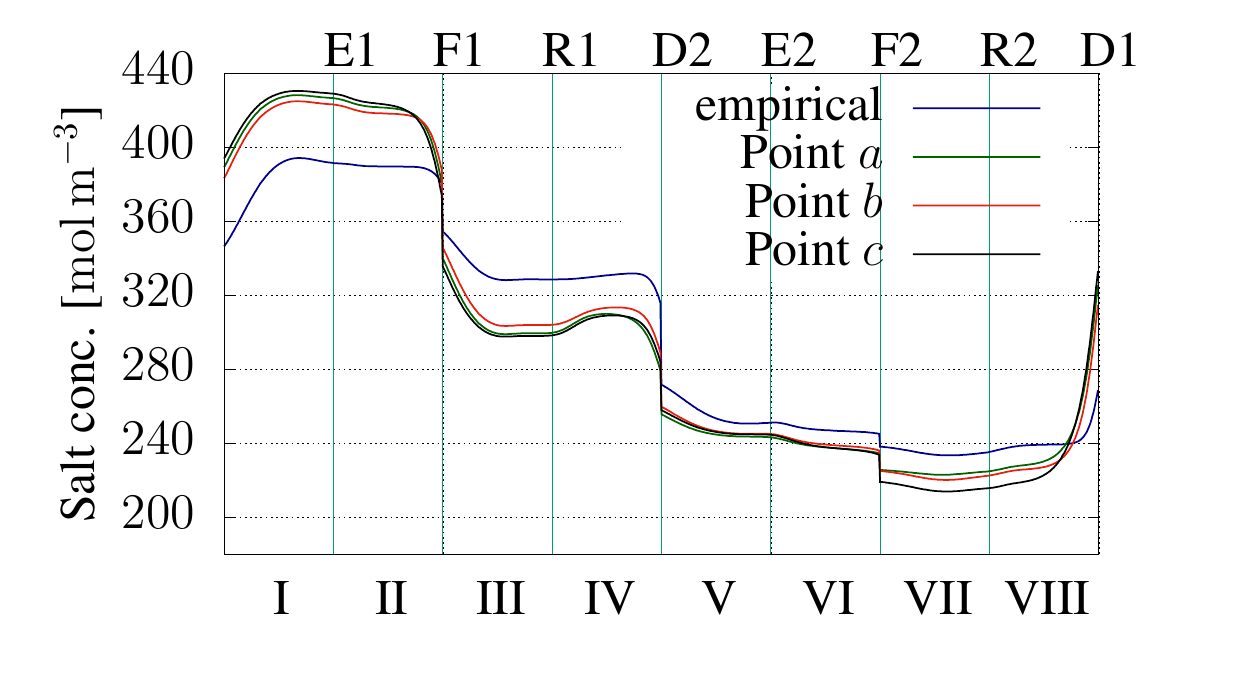}
    \caption{The actual salt profiles, $c_0(t=\tau, z)$, along the columns of the eight-zone scheme upon the \textsc{css} by controlling solvent gradients at inlet ports.}
    \label{fig:8-zone_salt}
\end{figure}

The salt profile, $c_0(t, z)$, along the eight-zone \textsc{iex-smb} unit upon \textsc{css}, $t = \tau$, is shown in Fig.~\ref{fig:8-zone_salt}.
The salt concentrations at inlet ports, $\sbk{c_0^{D1},\, c_0^{F1},\,\allowbreak c_0^{D2},\,\allowbreak c_0^{F2}}$, are $\sbk{442,\, 270,\,\allowbreak 240,\, 211} \si{\mole\per\cubic\metre}$ of the point $a$; $[440,\, 270,\,\allowbreak 240,\, 209]\, \si{\mole\per\cubic\metre}$ of the point $b$ and $[443,\, 272,\,\allowbreak 240,\, 217]\,\allowbreak \si{\mole\per\cubic\metre}$ of the point $c$.
Separating lyz at the port E1 is facilitated with the salt gap between zones I,II and zones III,IV; RNase and cyt are both weekly bounded in zones III,IV and to be separated in the second sub-unit.
Meanwhile, the bypass stream was diluted as described by Eq.~\ref{eq:dilute}.
As seen in Fig.~\ref{fig:8-zone_salt}, the salt concentration is decreased to ca.~\SI{240}{\mole\per\cubic\metre} in zones V,VI and \SI{220}{\mole\per\cubic\meter} in zones III,IV.
With these salt concentrations, RNase is weakly bounded and cyt is strongly bounded in the second sub-unit.
As seen from Fig.~\ref{fig:8-zone_salt}, the deviation of the salt profile along the columns of \textsc{iex-smb} is small, but the performance indicators significantly varied.
This indicates process design of \textsc{iex-smb} processes might be sensitive to operating conditions. 

Eight-zone schemes have fewer degrees of freedom than cascade schemes.
Hence, $\mathtt{Pu}_{\text{cyt}}^{E2}$ decreases from $99.81\%$ of the cascade scheme to $97.18\%$, based on the same yield, 0.95.
Moreover, in the cascade scheme, almost pure cyt can be collected, which can not be achieved with the eight-zone scheme.  
%However, the yield of the eight-zone scheme can be higher than that of the cascade scheme.
%This suggests that having more degrees of freedom in process design of \textsc{smb} might contribute to the purity performance.
However, the productivity of the eight-zone scheme can be higher than that of the cascade scheme.
The productivity of the cascade scheme, $\SI{3.13e-4}{\mole\per\cubic\metre\per\second}$ ($\mathcal{PI}_{\text{cyt}}^{E2} = [99.27\%,\, 0.96$]), is slightly lower than that of the eight-zone scheme, $\SI{3.25e-4}{\mole\per\cubic\metre\per\second}$ ($\mathcal{PI}_{\text{cyt}}^{E2} = [99.27\%,\, 0.97]$).
This suggests that the stationary phase is more efficiently utilized with integrated \textsc{smb} processes.

As no linear gradients can be applied in the \textsc{iex-smb} schemes, the eight-zone and cascade schemes have fewer degrees of freedom than multi-column continuous chromatography (\eg, \textsc{jo} and \textsc{mcsgp}).
Thus, we could conjecture that with proper process design, the \textsc{jo} and \textsc{mcsgp} would render more efficient and accurate elution of the model proteins.
However, this contribution is focused on the closed-loop \textsc{iex-smb} systems and the challenge of designing multi-step salt profiles by using modeling and high performance computing.

Three columns in each zone were chosen and used in both the cascade and eight-zone schemes.
In the empirical designs, neither one column nor two columns in each zone could lead to good separation performance. 
Having more columns in each zone and correspondingly shorter switching time can approximate the true moving bed chromatography.
However, introducing more columns into \textsc{iex-smb} system is computationally challenging and practically unrealistic.
Hence, three columns in each zone are a good compromise.

\subsection{Sensitivity of operating conditions}
\begin{figure}
    \centering
    \begin{subfigure}{0.7\textwidth}{\includegraphics[width=\textwidth]{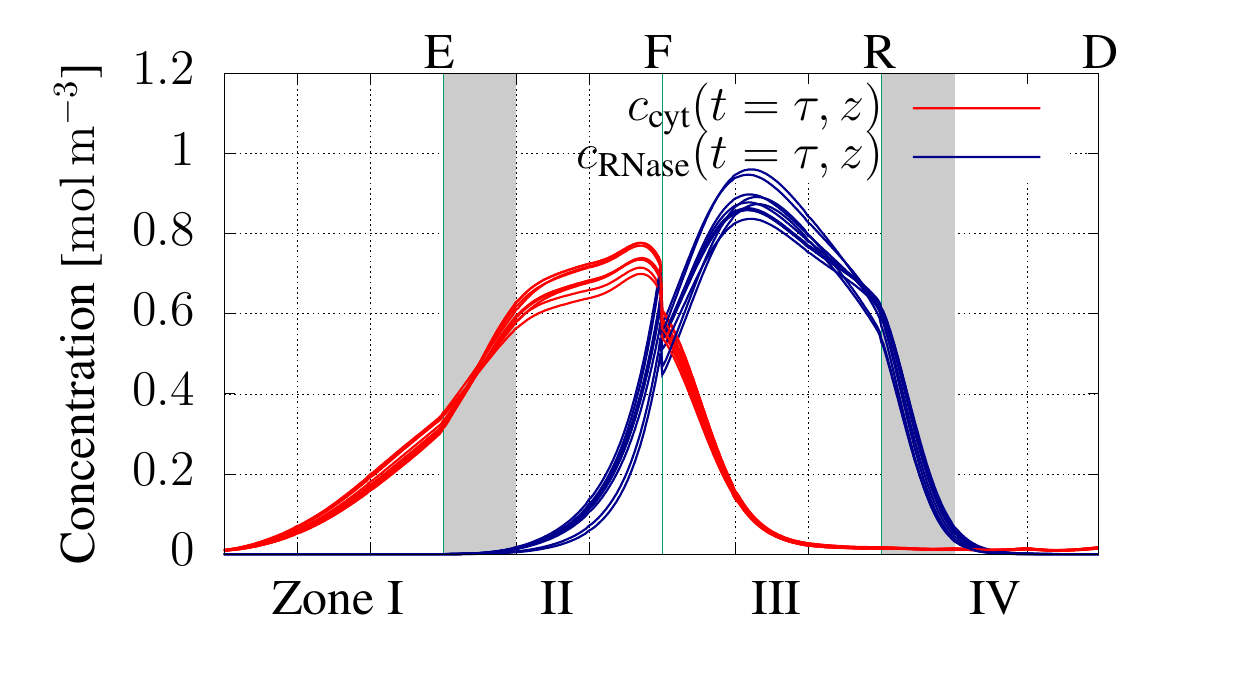}} \caption{Cascade scheme} \label{fig:dev_a} \end{subfigure}
    \begin{subfigure}{\textwidth}{\includegraphics[width=\textwidth]{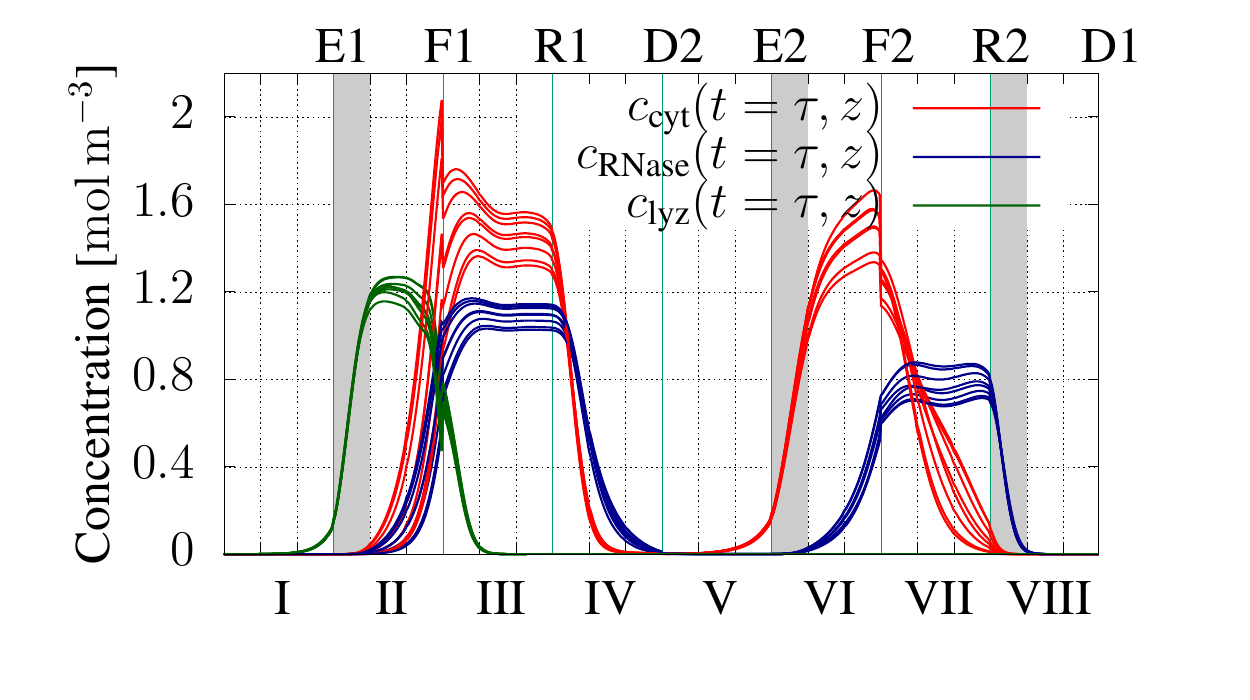}} \caption{Eight-zone scheme} \label{fig:dev_b} \end{subfigure}
    \caption{Perturbation of the \textsc{css} of eight-zone process with respect to the corresponding parameter distributions.}
    \label{fig:dev}
\end{figure}

As stated in the previous subsection, the \textsc{css} of both the cascade and the eight-zone schemes are sensitive to the operating conditions.
By utilizing \textsc{mcmc} sampling (cf.~\cite{he2019bayesian}), uncertainties of operating conditions are characterized by parameter distributions. % (\eg, Fig.~\ref{fig:post}; the other distributions are not shown).
%It illustrates that the operating conditions within the distribution deviations all render the same or similar performance indicators.
Posterior predictive checks were carried out to test the robustness of the two investigated processes.
Random parameter combinations from respective parameter distributions are used as arguments of forward simulations and to generate chromatograms.
The possible deviations of the investigated processes are illustrated with curve clusters, as shown in Fig.~\ref{fig:dev}.
%
%\begin{figure}
    %\centering
    %\includegraphics[width=0.5\textwidth]{ksdensity_8zone}
    %\caption{Schematic of a parameter distribution (\ie, salt concentration at the desorbent-I port) to characterize the parameter uncertainty.}
    %\label{fig:post}
%\end{figure}

As seen in Fig.~\ref{fig:dev}, the \textsc{css} are sensitive to the operating conditions in both configurations.
For instance, in the zones III, VI and VII of the eight-zone scheme, the curve deviations of the cyt and RNase components are quite significant; the same applies to the zones II and III of the cascade scheme.
However, in the outlet streams of the eight-zone configuration (\ie, E1, R2 and E2) and the cascade configuration (\ie, E and R), the deviations are tiny, which means perturbations in the operating conditions characterized by the parameter distributions do not essentially affect the withdrawn streams and consequentially the performance indicators.
Fig.~\ref{fig:dev} also indicates the eight-zone scheme is more robust than the cascade scheme.

Therefore, the studied \textsc{smb} configurations are robust provided the processes are performed within the perturbation tolerance.% as described by the Fig.~\ref{fig:post}.

\section{Conclusions}
%For macro-molecules, nonlinear binding models describe more accurate adsorption behaviours; but the nonlinearity of these binding models could make it impossible to use triangle theory to calculate operating parameters.
%In addition, triangle theory provdes no rules to design salt profiles in different zones of \textsc{iex-smb} processes.
We have presented model-based process designs of multi-step salt gradient single column batch chromatography and \textsc{iex-smb} chromatographic processes for separating ribonuclease, cytochrome and lysozyme on strong cation-exchanger SP Sepharose FF.
The general rate model with steric mass-action binding model was used in the column modeling.
Two network configurations of closed-loop \textsc{iex-smb} have been studied, \ie, one cascade scheme of two four-zone \textsc{iex-smb} sub-units and one integrated eight-zone scheme; the single column batch system has been studied for comparison.
The multi-objective sampling algorithm, Markov Chain Monte Carlo (\textsc{mcmc}), has been used to generate samples for approximating the Pareto optimal fronts.
This study shows that it is possible to design closed-loop \textsc{iex-smb} units for collecting the target, cytochrome, with high purity and yield.

In the numerical optimizations of \textsc{iex-smb}, initial values are required.
Empirical designs have been developed in this study to obtain the initial values, and the search domains of numerical optimization are based on the initial values with safety margins.
\textsc{mcmc} serves on the sampling purpose, which is interested in sampling Pareto optimal points as well as those near Pareto optimal.
With $\varepsilon$-constraint method of the multi-objective optimization, Pareto optimal fragments can be concerned to reduce expensive computational cost in large-scale \textsc{iex-smb} processes.
%Although in center-cut separations the other two components (\ie, ribonuclease and lysozyme) are treated as impurities, they can be collected with rather high performance indicators.

%comparison
The Pareto fronts show the information of the trade-off relationship between the conflicting indicators (\ie, purity and yield).
The cascade and eight-zone schemes have approximately the same performance, and both outperform the single column system.
In addition, the cascade scheme can produce almost pure target component with high yield, but lower robustness; the stationary phase is utilized more efficiently in the integrated eight-zone scheme in terms of productivity.
However, single column system does not undergo a time-consuming ramp-up phase before entering into a cyclic steady state, like it is in \textsc{smb} systems.
Starting a single column batch system prior to complex multi-column and \textsc{smb} systems is recommended.

% dilution
Dilution steps were applied in the bypass streams of the \textsc{iex-smb} schemes to reduce the salt concentrations, leading cytochrome to be more strongly bounded in the second sub-unit.
Without the dilution steps, the salt concentrations are too high such that ribonuclease and cytochrome are both weekly bounded to the stationary phase of the second sub-unit.
Though it is beneficial to implement dilution, it consumes more running buffer on one hand, and the cytochrome product needs to be more concentrated in the subsequent downstream processes on the other hand.

% outlook
There are many options for modeling mass transfer in a column, \eg, the equilibrium-dispersive model, the transport-dispersive model and the general rate model that can be combined with a multitude of binding models.
The software package used in this study is flexible in choosing between various modeling options for the columns and hold-up volumes.
\textsc{smb} processes do not necessarily need equivalent column numbers in each zone. 
Different column numbers in each zone can be applied, leading to mixed-integer programming.
The computational burden of \textsc{smb} processes in numerical optimizations can be reduced by optimizing initial states of the involved columns, or by using computationally cheap surrogate models.
Multi-dimensional Pareto optimal fronts can also be further investigated, taking buffer consumption, productivity or processing time into account.

\section{Acknowledgments}
This work is partially supported by the China Scholarship Council (CSC, no.~201408310124) and partially by the National Key Research and Development Program of China (NKRDP, grant.~2017YFB0309302).

\bibliographystyle{model5-names}

\bibliography{references}

\end{document}